\pdfoutput=1
\documentclass[11pt,a4paper]{article}
\usepackage{jheppub}
\usepackage[caption=false]{subfig}
\usepackage{graphicx}
\usepackage{empheq}
\usepackage{hyperref}
\usepackage{multirow}
\usepackage{enumerate}
\usepackage{color}

\newcommand{\beq}{\begin{eqnarray}}
\newcommand{\eeq}{\end{eqnarray}}
\newcommand{\non}{\nonumber\\}

\newcommand{\p}{\partial}

\renewcommand{\d}[1]{\ensuremath{\operatorname{d}\!{#1}}}

\newcommand{\red}[1]{{\color{red}#1}}
\renewcommand{\i}{\mbox{i}}
\newcommand{\e}{\mbox{e}}

\begin{document}

\title{Boundary charges and integral identities for solitons in
$(d+1)$-dimensional field theories}

\author{Sven Bjarke Gudnason,${}^{1}$}
\author{Zhifeng Gao${}^2$ and}
\author{Yisong Yang${}^3$}
\affiliation{${}^1$Institute of Modern Physics, Chinese Academy of Sciences,
  Lanzhou 730000, China}
\affiliation{${}^2$Institute of Contemporary Mathematics, School of
Mathematics, Henan University, Kaifeng, Henan 475004, China} 
\affiliation{${}^3$Courant Institute of Mathematical Sciences, New
York University, New York, NY 10012, USA} 
\emailAdd{bjarke(at)impcas.ac.cn}
\emailAdd{gzf(at)henu.edu.cn}
\emailAdd{yisongyang(at)nyu.edu}

\abstract{
We establish a 3-parameter family of integral identities to be used on
a class of theories possessing solitons with spherical symmetry in $d$
spatial dimensions.
The construction provides five boundary charges that are related to
certain integrals of the profile functions of the solitons in
question.
The framework is quite generic and we give examples of both
topological defects (like vortices and monopoles) and topological
textures (like Skyrmions) in 2 and 3 dimensions.
The class of theories considered here is based on a kinetic term
and three functionals often encountered in reduced Lagrangians for
solitons. 
One particularly interesting case provides a generalization of the
well-known Pohozaev identity.
Our construction, however, is fundamentally different from scaling
arguments behind Derrick's theorem and virial relations.
For BPS vortices, we find interestingly an infinity of integrals
simply related to the topological winding number.
}

%\keywords{Solitons, boundary charges, integral identities} 

\maketitle

%%%%%%%%%%%%%%%%%%%%%%%%%%%%%%%%%%%%%%%%%%%%%%%%%%%%%%%%%%%

\section{Introduction}

Solitons play an important role in non-perturbative physics of all
kinds.
Except for a few special cases, such as self-dual instantons or
dimensional reductions thereof, solitonic systems are often not
integrable in more than one codimension.
Vortices at critical coupling are described by the Taubes
equation \cite{Taubes:1979tm} which is not integrable.
The exception, however, is the case where they are placed on a
hyperbolic plane of a particular constant curvature and the Taubes
equation is modified into the Liouville equation and hence becomes 
integrable \cite{Witten:1976ck}. This case corresponds, 
however, to the self-dual instanton equation on the background
$S^2\times\mathbb{H}^2$ and thus it is not surprising that it is
integrable. 

Hence most soliton solutions can only be obtained by means of
numerical techniques and in case of numerically challenging parts of
the parameter space, mathematical identities other than the equation
of motion can be useful to check the validity and precision of the
obtained numerical solutions, because the former often involve
lower-order derivatives.

For simplicity, we focus on a simple class of theories with spherical
symmetry, that consists of a single profile function, $f$, which has a
Lagrangian description that involves a standard kinetic term, two
functionals of $f$: one purely dependent on the field and one that
also allows for the coordinate ($r$), and finally a modified kinetic
term, that is simply parametrized by a functional of $f$ and $r$.
The potential term that allows for a dependence on $r$ can be thought
of as a standard kinetic term, where the derivatives act on the angular
parts of the underlying field configuration; for sigma model fields,
this looks in the reduced Lagrangian like an $r$-dependent potential. 
Since we will be interested in topological solitons, we will not
consider time dependence; this makes our results applicable for both
relativistic as well as non-relativistic theories. 
The reason for imposing spherical symmetry is simply to reduce the
equation of motion to a one-dimensional one.
We do, however, keep the number of spatial dimensions general, so that
spherically symmetric systems in $d$ dimensions are treated on equal
footing. 

In practice our construction is based on the equation of motion along
the radial coordinate, multiplied by an appropriately chosen weight
function, that allows us to turn the standard ``conjugate
momentum''\footnote{Strictly speaking, this is not a momentum, but a
variation with respect to the radial derivative of the field profile
$f$. },
$\frac{\p\mathcal{L}}{\p f'}$, into a generalized function which
remains as a boundary term (total derivative) in the equation.
Once we integrate the equation, we get the first boundary charge,
which we shall denote as the type-{\bf 0} boundary charge. 

A more interesting outcome of our construction of identities than
simply making relations that can be used for checking numerical
solutions, is that -- in the class of theories that we are considering
-- we find \emph{five} different types of boundary terms, yielding the 
possibility of picking up what we call boundary charges.
The boundary charge consists of some combination of solution
parameters, Lagrangian parameters, coupling constants or boundary
conditions -- evaluated at the endpoint(s) of the integration range.
We would like to promote the boundary charges to be thought of as
topological boundary charges, although we do not have a mathematical
understanding (definition) of them yet.
There are two reasons for this promotion; the first is due to the fact
that any angular dependence in the higher-dimensional soliton will
often contribute the winding number or topological degree to exactly
these boundary charges. The second reason is the similarity of the
boundary charges and so-called domain wall topological charges
(although in the case of the domain wall, it is not a difference of
the potential, but of the superpotential, evaluated on the endpoints
of the field values). 

After setting up the general framework of identities and illustrating
various flavors or simplifications of the main identity, we consider
some examples of solitons in 2 and 3 dimensions.\footnote{Since most
one-dimensional solitons are exactly solvable (integrable), we will
focus on $d>1$ dimensional solitons.}
We obtain many interesting relations; in particular, we find a
generalization of the Pohozaev identity \cite{Pohozaev:1965}, see
e.g.~Eq.~\eqref{eq:idVI+VIII_moment} below.
The generalization picks up three different boundary charges, that we
call type {\bf 0}, type {\bf IIa} and type {\bf IIb}, respectively,
and relates them to the potential energy of the system as well as
other terms that are generically non-vanishing with respect to Derrick
scaling \cite{Derrick:1964ww} (that means classically non-conformal). 

The most general form of our identity is a 3-parameter family of
relations that nontrivially provide different integrals of the
fields.
The three parameters can be used to pick up the desired boundary
charges of the theory at hand (if possible), but also it can be used
to avoid unwanted divergences that often plague global solitons (as a
consequence of Derrick's theorem). 

\begin{figure}[!t]
\begin{center}
\includegraphics[width=0.6\linewidth]{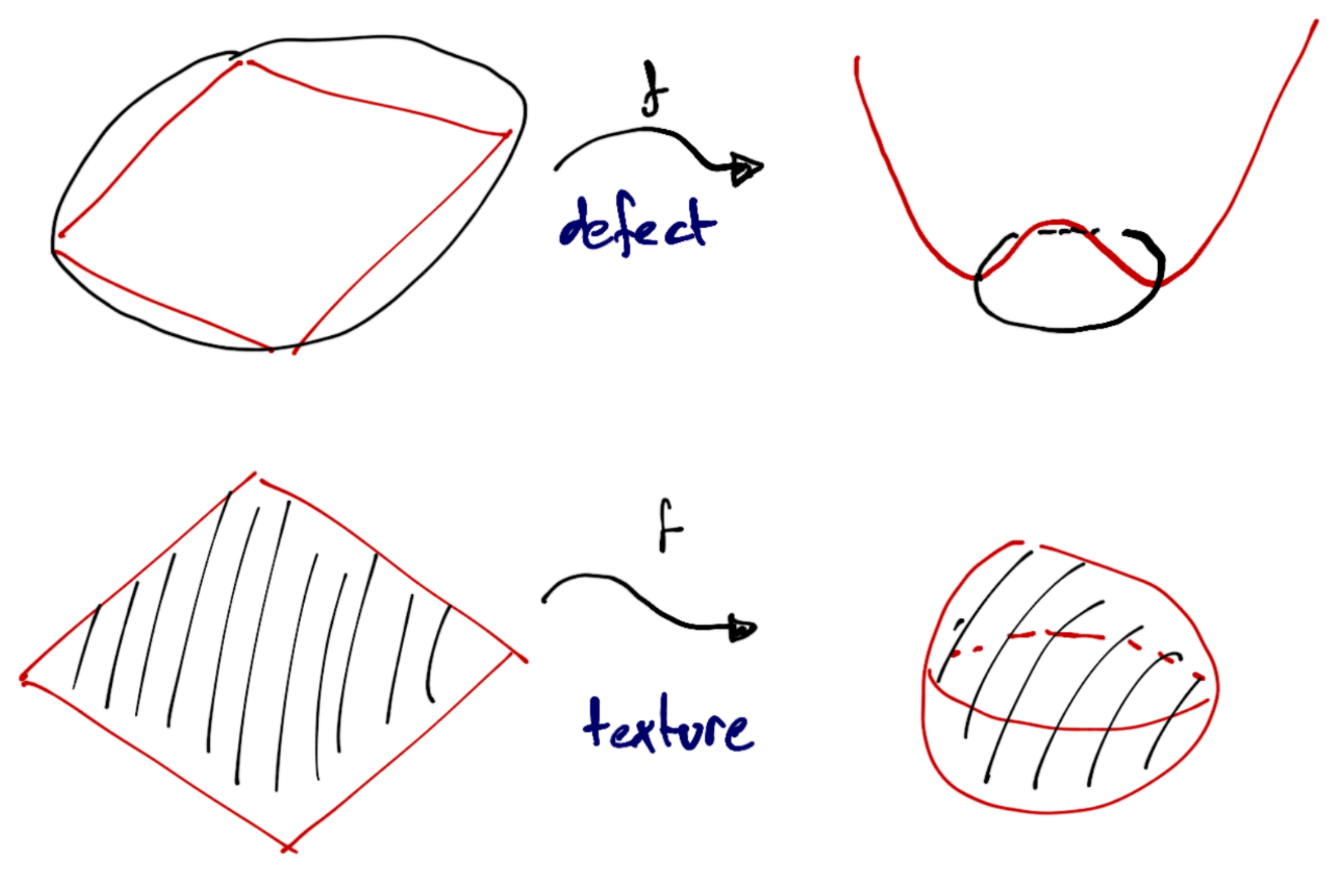}
\caption{Topological defects versus topological textures.}
\label{fig:defectsandtextures}
\end{center}
\end{figure}

Since the boundary charges crucially depend on the solutions and in
turn on their boundary conditions, it will prove useful to define the
following notation for topological solitons (which we will refer to
throughout the paper): we distinguish between topological defects and
topological textures, see Fig.~\ref{fig:defectsandtextures}.
The topological defect in $d$ spatial dimensions, is defined by having
a nontrivial $(d-1)$-th homotopy group of the target space of the
theory 
\beq
\pi_{d-1}(G/H),
\eeq
because the soliton solution is a map from the boundary of space
$\p\mathbb{R}^{d}\simeq S^{d-1}$ to $G/H$, where $G$ is an internal
symmetry group and $H$ is a stabilizer.
The prime example is a vortex in $d=2$ dimensions in a theory where an
internal U(1) symmetry is completely broken
\beq
\pi_1(S^1) = \mathbb{Z},
\eeq
which constitutes the group of integers.

The topological texture, on the other hand, is by construction a map
of the whole space with infinity compactified to a point,
$\mathbb{R}^d\cup\{\infty\}\simeq S^d$ to the target space $G/H$.
This yields the homotopy group for a topological texture
\beq
\pi_d(G/H),
\eeq
which is clearly different from that of a defect. 

For BPS vortices of both standard Abelian-Higgs type and 
Chern-Simons type, we find type-{\bf 0} boundary charges proportional
to the winding number, due to the logarithmic singularity at the
origin of the auxiliary profile function. 
We find what we later will define as type-{\bf II} boundary charges
for global vortices, the Skyrme vortices and the global monopoles, but
not for the topological textures (such as the baby Skyrmions and the
Skyrmions).
On the other hand, we find type-{\bf I} boundary charges for the
baby Skyrmions, the Skyrme vortices and the Skyrmions, viz.~only for
theories with higher-order derivative terms (for the definition of the
types of boundary charges, see the next section). 

%% global vortex: IIa + IIb
%% baby Skyrmions: Ia
%% skyrme vortex: Ia + IIa
%% global monopoles: IIa
%% Skyrmions: Ia

A comment in store is about the origin of the boundary charges.
We would like to stress that naive scaling arguments may miss the
boundary charges and hence, in certain cases with nontrivial boundary 
conditions or topological winding at infinity, will get the wrong
answer.
Therefore, our relations are not simply based on scaling arguments,
like Derrick's theorem or virial theorems.
Our identities are also not directly related to the Noether theorem.
As mentioned above, the exact mathematical origin of the boundary
charges are not yet known.

Although the relations we find are interesting themselves, we flesh out
many details of the framework of finding the boundary charges in the
hope that the interested reader will better understand the mechanisms
at work and perhaps further generalize the framework.

The paper is organized as follows.
In Sec.~\ref{sec:identity} we derive the identity and illustrate the
cases giving rise to the five boundary charges as well as many cases
that simplify the main identity for our class of theories.
Secs.~\ref{sec:example2} and \ref{sec:example3} illustrate the
framework of identities with examples of topological solitons of both
defect type and texture type in two and three dimensions,
respectively. 
Finally, Sec.~\ref{sec:discussion} concludes the paper with a
discussion and outlook on how the results can be generalized.
Some selected numerical checks are delegated to
appendix \ref{app:numerical_checks}.

\section{The identity}\label{sec:identity}

Let us consider the static Lagrangian density in $d$ spatial
dimensions (i.e~on $\mathbb{R}^d$) of the form 
\beq
\mathcal{L}[f,f',r] =
-\frac{1}{2} f^{'2}
- F[f,r]
- \frac{1}{2}G[f,r]f^{'2}
- V[f],
\label{eq:Lcankin}
\eeq
where $f=f(r)$ is a radial profile function and
$f'\equiv\frac{\d{f}}{\d{r}}$.
The Euler-Lagrange equation associated with the action
$\int_0^\infty \d{r}\; r^{d-1}\mathcal{L}$ reads
\beq
\frac{1}{r^{d-1}}\frac{\d{}}{\d{r}} \left(r^{d-1}(1+G)f'\right)
- \frac{\p F}{\p f}
- \frac{1}{2} \frac{\p G}{\p f} f^{'2}
- \frac{\p V}{\p f} = 0.
\eeq
The basic construction of our class of identities is to multiply the
equation of motion by $r^{(d-1)\kappa} f^{'\mu-1} (1+G)^{\nu-1}$ and
rearrange the terms into total derivatives where
possible\footnote{Note that we distinguish between the total
derivative and the partial derivative, e.g.~$\frac{\d{G}}{\d{r}}=\frac{\p
G}{\p f}f'+\frac{\p G}{\p r}$.}
\begin{align}
&\frac{1}{\lambda}\frac{\d{}}{\d{r}}
\left(r^{(d-1)\kappa} (1+G)^\nu f^{'\mu}\right) 
+(d-1)\left(1-\frac{\kappa}{\lambda}\right) r^{(d-1)\kappa-1} (1+G)^\nu
  f^{'\mu} \non
&+\left(1-\frac{\mu}{\lambda}\right)r^{(d-1)\kappa}(1+G)^\nu
  f^{'\mu-1}f'' 
+\left(\frac{1}{2}-\frac{\nu}{\lambda}\right) r^{(d-1)\kappa}
  (1+G)^{\nu-1} \frac{\p G}{\p f} f^{'\mu+1} \non
&+\left(1-\frac{\nu}{\lambda}\right) r^{(d-1)\kappa}
  (1+G)^{\nu-1} \frac{\p G}{\p r} f^{'\mu} 
-r^{(d-1)\kappa} \frac{\p F}{\p f} (1+G)^{\nu-1} f^{'\mu-1} \non
&-r^{(d-1)\kappa} (1+G)^{\nu-1} \frac{\p V}{\p f} f^{'\mu-1} = 0.
\label{eq:id_diff}
\end{align}
In the above equation, we formally take all parameters such that all
terms are real and well-defined quantities.
The first term, proportional to $1/\lambda$ is added as a
total derivative and subtracted again as integrals of the derivative
acting on each factor.
$\lambda$ should be chosen to be equal to one of the other three
parameters ($\kappa$, $\mu$ or $\nu$), see below.
Integrating the above equation with respect to $r$ yields
\begin{align}
&\left.\frac{1}{\lambda}
  r^{(d-1)\kappa} (1+G)^\nu f^{'\mu}\right|_0^\infty
+(d-1)\left(1-\frac{\kappa}{\lambda}\right)
  \int_0^\infty\d{r}\; r^{(d-1)\kappa-1} (1+G)^\nu f^{'\mu} \non
&+\left(1-\frac{\mu}{\lambda}\right)
  \int_0^\infty\d{r}\; r^{(d-1)\kappa}(1+G)^\nu f^{'\mu-1}f'' 
+\left(\frac{1}{2}-\frac{\nu}{\lambda}\right)
  \int_0^\infty\d{r}\; r^{(d-1)\kappa}(1+G)^{\nu-1} \frac{\p G}{\p f}
  f^{'\mu+1} \non 
&+\left(1-\frac{\nu}{\lambda}\right)
  \int_0^\infty\d{r}\; r^{(d-1)\kappa} (1+G)^{\nu-1} \frac{\p G}{\p r}
  f^{'\mu}  
-\int_0^\infty\d{r}\; r^{(d-1)\kappa} \frac{\p F}{\p f} (1+G)^{\nu-1}
  f^{'\mu-1} \non 
&-\int_0^\infty\d{r}\; r^{(d-1)\kappa} (1+G)^{\nu-1} \frac{\p V}{\p f}
  f^{'\mu-1} = 0. 
\label{eq:id}
\end{align}
which is the basic identity and $\kappa,\mu,\nu$ and $\lambda$ are
real (not necessarily integer) parameters.
Although it is very complicated in its most general form, we will see 
that certain choices for $\kappa,\mu,\nu$ and $\lambda$, will give
interesting subclasses of integral identities depending on the theory
under study.

The first boundary term is found by integrating the total derivative
in Eq.~\eqref{eq:id_diff}; we will call this the boundary charge of
type {\bf 0}.

Keeping $\lambda$ as a parameter is the most general case.
One can always integrate the third term by parts to get rid
of $f''$; that is, however, exactly the same as setting $\lambda=\mu$,
which also eliminates $f''$ from the identity.
Not having to know $f''$ can be advantageous numerically, because it
is often easier to calculate $f'$ than $f''$.
$\lambda$ is not a genuine parameter on the same footing as $\kappa$,
$\mu$ and $\nu$, because the terms that $1/\lambda$ multiplies add up
to zero. The purpose of $\lambda$ is to leave the choice of setting
$\lambda=\mu$, $\lambda=\kappa$ or $\lambda=\nu$ open and $\lambda$ is
thus left as a free parameter to make a unified framework of
identities.  

In the construction, we have assumed a standard kinetic term
($-\tfrac{1}{2}f'^2$) in the Lagrangian density. If such term is not
present, a simple substitution can be made to get the appropriate
identity: $(1+G)\to G$. 

Let us discuss different possibilities where the general
identity \eqref{eq:id} simplifies:
\begin{enumerate}[I]
\item\label{id:I} $\lambda=\kappa$ makes the second term vanish,
\item\label{id:II} $\lambda=\mu$ makes the third term vanish,
\item\label{id:III} $\lambda=2\nu$ eliminates the fourth term; if $G$
is a function of only $f$, this may be useful,
\item\label{id:IV} $\lambda=\nu$ eliminates the fifth term; if $G$ is
a function of only $r$, this may be useful,
\item\label{id:V} $G=0$ simplifies all terms and eliminates the
second and third terms,
\item\label{id:VI} if $G\neq 0$, then $\nu=1$ simplifies the $(1+G)$
factors in the last four terms,
\item\label{id:VII} $\mu=1$ allows for integration by parts of the
second and fifth terms,
\item\label{id:VIII} $\mu=2$ allows for the possibility to rewrite the
last (two) term(s) into an integral over $V$ (and $F$); this is
desirable if one wants to determine the potential energy.
\end{enumerate}
Let us write out the simplified identities in turn, starting with the
identity \ref{id:I} ($\lambda=\kappa$):
\begin{align}
&\left.\frac{1}{\kappa}
  r^{(d-1)\kappa} (1+G)^\nu f^{'\mu}\right|_0^\infty
+\left(1-\frac{\mu}{\kappa}\right)
  \int_0^\infty\d{r}\; r^{(d-1)\kappa}(1+G)^\nu f^{'\mu-1}f'' \non
&+\left(\frac{1}{2}-\frac{\nu}{\kappa}\right)
  \int_0^\infty\d{r}\; r^{(d-1)\kappa}(1+G)^{\nu-1} \frac{\p G}{\p f}
  f^{'\mu+1} \non
&+\left(1-\frac{\nu}{\kappa}\right)
  \int_0^\infty\d{r}\; r^{(d-1)\kappa} (1+G)^{\nu-1} \frac{\p G}{\p r}
  f^{'\mu} 
-\int_0^\infty\d{r}\; r^{(d-1)\kappa} \frac{\p F}{\p f} (1+G)^{\nu-1}
  f^{'\mu-1} \non 
&-\int_0^\infty\d{r}\; r^{(d-1)\kappa} (1+G)^{\nu-1} \frac{\p V}{\p f}
  f^{'\mu-1} = 0. 
\label{eq:idI}
\end{align}
Identity \ref{id:II} ($\lambda=\mu$):
\begin{align}
&\left.\frac{1}{\mu}
  r^{(d-1)\kappa} (1+G)^\nu f^{'\mu}\right|_0^\infty
+(d-1)\left(1-\frac{\kappa}{\mu}\right)
  \int_0^\infty\d{r}\; r^{(d-1)\kappa-1} (1+G)^\nu f^{'\mu} \non
&+\left(\frac{1}{2}-\frac{\nu}{\mu}\right)
  \int_0^\infty\d{r}\; r^{(d-1)\kappa}(1+G)^{\nu-1} \frac{\p G}{\p f}
  f^{'\mu+1} \non
&+\left(1-\frac{\nu}{\mu}\right)
  \int_0^\infty\d{r}\; r^{(d-1)\kappa} (1+G)^{\nu-1} \frac{\p G}{\p r}
  f^{'\mu}  
-\int_0^\infty\d{r}\; r^{(d-1)\kappa} \frac{\p F}{\p f} (1+G)^{\nu-1}
  f^{'\mu-1} \non 
&-\int_0^\infty\d{r}\; r^{(d-1)\kappa} (1+G)^{\nu-1} \frac{\p V}{\p f}
  f^{'\mu-1} = 0. 
\label{eq:idII}
\end{align}
Identity \ref{id:III} ($\lambda=2\nu$):
\begin{align}
&\left.\frac{1}{2\nu}
  r^{(d-1)\kappa} (1+G)^\nu f^{'\mu}\right|_0^\infty
+(d-1)\left(1-\frac{\kappa}{2\nu}\right)
  \int_0^\infty\d{r}\; r^{(d-1)\kappa-1} (1+G)^\nu f^{'\mu} \non
&+\left(1-\frac{\mu}{2\nu}\right)
  \int_0^\infty\d{r}\; r^{(d-1)\kappa}(1+G)^\nu f^{'\mu-1}f'' 
+\frac{1}{2}\int_0^\infty\d{r}\; r^{(d-1)\kappa} (1+G)^{\nu-1}
  \frac{\p G}{\p r} f^{'\mu} \non
&-\int_0^\infty\d{r}\; r^{(d-1)\kappa} \frac{\p F}{\p f} (1+G)^{\nu-1}
  f^{'\mu-1} 
-\int_0^\infty\d{r}\; r^{(d-1)\kappa} (1+G)^{\nu-1} \frac{\p V}{\p f}
  f^{'\mu-1} = 0. 
\label{eq:idIII}
\end{align}
Identity \ref{id:IV} ($\lambda=\nu$):
\begin{align}
&\left.\frac{1}{\nu}
  r^{(d-1)\kappa} (1+G)^\nu f^{'\mu}\right|_0^\infty
+(d-1)\left(1-\frac{\kappa}{\nu}\right)
  \int_0^\infty\d{r}\; r^{(d-1)\kappa-1} (1+G)^\nu f^{'\mu} \non
&+\left(1-\frac{\mu}{\nu}\right)
  \int_0^\infty\d{r}\; r^{(d-1)\kappa}(1+G)^\nu f^{'\mu-1}f'' 
-\frac{1}{2}\int_0^\infty\d{r}\; r^{(d-1)\kappa}(1+G)^{\nu-1}
  \frac{\p G}{\p f} f^{'\mu+1} \non 
&-\int_0^\infty\d{r}\; r^{(d-1)\kappa} \frac{\p F}{\p f} (1+G)^{\nu-1}
  f^{'\mu-1}
-\int_0^\infty\d{r}\; r^{(d-1)\kappa} (1+G)^{\nu-1} \frac{\p V}{\p f}
  f^{'\mu-1} = 0. 
\label{eq:idIV}
\end{align}
Identity \ref{id:V} ($G=0$):
\begin{align}
&\left.\frac{1}{\lambda}
  r^{(d-1)\kappa} f^{'\mu}\right|_0^\infty
+(d-1)\left(1-\frac{\kappa}{\lambda}\right)
  \int_0^\infty\d{r}\; r^{(d-1)\kappa-1} f^{'\mu} 
+\left(1-\frac{\mu}{\lambda}\right)
  \int_0^\infty\d{r}\; r^{(d-1)\kappa} f^{'\mu-1}f'' \non
&-\int_0^\infty\d{r}\; r^{(d-1)\kappa} \frac{\p F}{\p f}
  f^{'\mu-1} 
-\int_0^\infty\d{r}\; r^{(d-1)\kappa} \frac{\p V}{\p f}
  f^{'\mu-1} = 0. 
\label{eq:idV}
\end{align}
Identity \ref{id:VI} ($\nu=1$):
\begin{align}
&\left.\frac{1}{\lambda}
  r^{(d-1)\kappa} (1+G) f^{'\mu}\right|_0^\infty
+(d-1)\left(1-\frac{\kappa}{\lambda}\right)
  \int_0^\infty\d{r}\; r^{(d-1)\kappa-1} (1+G) f^{'\mu} \non
&+\left(1-\frac{\mu}{\lambda}\right)
  \int_0^\infty\d{r}\; r^{(d-1)\kappa}(1+G) f^{'\mu-1}f'' 
+\left(\frac{1}{2}-\frac{1}{\lambda}\right)
  \int_0^\infty\d{r}\; r^{(d-1)\kappa} \frac{\p G}{\p f} f^{'\mu+1} \non 
&+\left(1-\frac{1}{\lambda}\right)
  \int_0^\infty\d{r}\; r^{(d-1)\kappa} \frac{\p G}{\p r} f^{'\mu}  
-\int_0^\infty\d{r}\; r^{(d-1)\kappa} \frac{\p F}{\p f} f^{'\mu-1} \non
&-\int_0^\infty\d{r}\; r^{(d-1)\kappa} \frac{\p V}{\p f} f^{'\mu-1} = 0. 
\label{eq:idVI}
\end{align}
Identity \ref{id:VII} ($\mu=1$):
\begin{align}
&\left.\frac{1}{\lambda}
  r^{(d-1)\kappa} (1+G)^\nu f'\right|_0^\infty 
+\left.(d-1)\left(1-\frac{\kappa}{\lambda}\right)
  r^{(d-1)\kappa-1} (1+G)^\nu f\right|_0^\infty \non
&+\left.\left(1-\frac{\nu}{\lambda}\right)
  r^{(d-1)\kappa} (1+G)^{\nu-1} \frac{\p G}{\p r}
  f\right|_0^\infty \non
&-(d-1)\left(1-\frac{\kappa}{\lambda}\right)\left[(d-1)\kappa-1\right]
  \int_0^\infty \d{r}\; r^{(d-1)\kappa-2} (1+G)^\nu f \non
&-(d-1)\left(1-\frac{\kappa}{\lambda}\right)\nu
  \int_0^\infty \d{r}\; r^{(d-1)\kappa-1} (1+G)^{\nu-1}
  \frac{\p G}{\p f} f f' \non
&+\left(1-\frac{1}{\lambda}\right)
  \int_0^\infty\d{r}\; r^{(d-1)\kappa}(1+G)^\nu f'' 
+\left(\frac{1}{2}-\frac{\nu}{\lambda}\right)
  \int_0^\infty\d{r}\; r^{(d-1)\kappa}(1+G)^{\nu-1} \frac{\p G}{\p f}
  f^{'2} \non
&-(d-1)\left(\kappa+\nu-\frac{2\kappa\nu}{\lambda}\right)
  \int_0^\infty \d{r}\; r^{(d-1)\kappa-1} (1+G)^{\nu-1}
  \frac{\p G}{\p r} f \non
&-\left(1-\frac{\nu}{\lambda}\right)(\nu-1)
  \int_0^\infty \d{r}\; r^{(d-1)\kappa} (1+G)^{\nu-2}
  \frac{\p G}{\p r}
  \left(\frac{\p G}{\p f} f' + \frac{\p G}{\p r}\right) f \non
&-\left(1-\frac{\nu}{\lambda}\right) \int_0^\infty \d{r}\;
  r^{(d-1)\kappa} (1+G)^{\nu-1}
  \left(\frac{\p^2G}{\p f\p r}f' + \frac{\p^2 G}{\p r^2}\right) f 
-\int_0^\infty\d{r}\; r^{(d-1)\kappa} \frac{\p F}{\p f} (1+G)^{\nu-1} \non
&-\int_0^\infty\d{r}\; r^{(d-1)\kappa} (1+G)^{\nu-1} \frac{\p V}{\p f} = 0. 
\label{eq:idVII}
\end{align}
Identity \ref{id:VIII} ($\mu=2$):
\begin{align}
&\left.\frac{1}{\lambda}
  r^{(d-1)\kappa} (1+G)^\nu f^{'2}\right|_0^\infty
\left.\mathop-r^{(d-1)\kappa} F (1+G)^{\nu-1}\right|_0^\infty
\left.\mathop-r^{(d-1)\kappa}(1+G)^{\nu-1} V\right|_0^\infty 
\non
&+(d-1)\left(1-\frac{\kappa}{\lambda}\right)
  \int_0^\infty\d{r}\; r^{(d-1)\kappa-1} (1+G)^\nu f^{'2} 
+\left(1-\frac{2}{\lambda}\right)
  \int_0^\infty\d{r}\; r^{(d-1)\kappa}(1+G)^\nu f' f'' \non
&+\left(\frac{1}{2}-\frac{\nu}{\lambda}\right)
  \int_0^\infty\d{r}\; r^{(d-1)\kappa}(1+G)^{\nu-1} \frac{\p G}{\p f}
  f^{'3} 
+\left(1-\frac{\nu}{\lambda}\right)
  \int_0^\infty\d{r}\; r^{(d-1)\kappa} (1+G)^{\nu-1} \frac{\p G}{\p r}
  f^{'2} \non
&+\int_0^\infty \d{r}\; r^{(d-1)\kappa} \frac{\p F}{\p r}
  (1+G)^{\nu-1} 
+(d-1)\kappa \int_0^\infty \d{r}\; r^{(d-1)\kappa-1} F
  (1+G)^{\nu-1} \non
&+(\nu-1)\int_0^\infty \d{r}\; r^{(d-1)\kappa} F (1+G)^{\nu-2}
  \left(\frac{\p G}{\p f}f' + \frac{\p G}{\p r}\right) \non
&+(d-1)\kappa \int_0^\infty \d{r}\; r^{(d-1)\kappa-1} (1+G)^{\nu-1} V \non
&+(\nu-1) \int_0^\infty \d{r}\; r^{(d-1)\kappa} (1+G)^{\nu-2}
  \left(\frac{\p G}{\p f}f' + \frac{\p G}{\p r}\right) V
=0.
\label{eq:idVIII}
\end{align}
Of course many combinations of the above six simplifications can be
made, depending on the case at hand.

The last two cases also provide boundary charges, by integrating by
parts.
The $\mu=1$ (identity \ref{id:VII}) case gives two new boundary
charges: the first (second term in Eq.~\eqref{eq:idVII}) is due to
the derivative in the equation of motion acting on the $r^{d-1}$
factor of the volume form; this term is sometimes called the
centrifugal term. 
We will denote this the type {\bf Ia} boundary charge. 
The next boundary term (third term in Eq.~\eqref{eq:idVII}) comes from
integrating by parts the term due to the (partial) radial derivative
acting on the function $G$. This happens only to $G$ because $G$ is
the prefactor of $f^{'2}$ in the Lagrangian \eqref{eq:Lcankin}. 
We will denote this as the type {\bf Ib} boundary charge.

The next case giving boundary terms is the case of $\mu=2$
(identity \ref{id:VIII}), again yielding two new boundary terms:
the first boundary term (second term in Eq.~\eqref{eq:idVIII}) is due
to the variation of $F$ with respect to the profile function $f$; once
it is multiplied by $f'$, it can be integrated by parts yielding a
boundary charge involving $F$ itself.
We denote this the type {\bf IIa} boundary charge.
Finally, there is the last boundary term (third term in
Eq.~\eqref{eq:idVIII}), which is exactly as the latter, except for
$V$. Once integrated by parts the boundary charge depends on $V$ (as 
opposed to the variation of $V$).
We denote this the type {\bf IIb} boundary charge. 

Let us illustrate a few combinations of identities, as useful
examples:\\ 
Identity \ref{id:II}+\ref{id:V} ($\lambda=\mu$ and $G=0$):
\begin{align}
&\left.\frac{1}{\mu}
  \left(r^{(d-1)\kappa} f^{'\mu}\right)\right|_0^\infty
+(d-1)\left(1-\frac{\kappa}{\mu}\right)
  \int_0^\infty\d{r}\; r^{(d-1)\kappa-1} f^{'\mu} 
-\int_0^\infty\d{r}\; r^{(d-1)\kappa} \frac{\p F}{\p f}
  f^{'\mu-1} \non
&-\int_0^\infty\d{r}\; r^{(d-1)\kappa} \frac{\p V}{\p f}
  f^{'\mu-1} = 0. 
\label{eq:idII+V}
\end{align}
This combination is useful for theories with $G=0$ where one does not
want to include $f''$.

Identity \ref{id:I}+\ref{id:II}+\ref{id:V} ($\lambda=\kappa=\mu$ and
$G=0$): 
\begin{align}
&\left.\frac{1}{\kappa} r^{(d-1)\kappa} f^{'\kappa}\right|_0^\infty
-\int_0^\infty\d{r}\; r^{(d-1)\kappa} \frac{\p F}{\p f}
  f^{'\kappa-1} 
-\int_0^\infty\d{r}\; r^{(d-1)\kappa} \frac{\p V}{\p f}
  f^{'\kappa-1} = 0. 
\label{eq:idI+II+V}
\end{align}
This combination is particularly useful for auxiliary Lagrangians
reproducing BPS solitons because the boundary term (the first term)
can pick up powers of the topological number $N$ of the soliton
(i.e.~a type-{\bf 0} boundary charge) and is 
simply related to integrals of $\frac{\p F}{\p f}$ and
$\frac{\p V}{\p f}$.
See the examples in Secs.~\ref{sec:AHBPSv} and \ref{sec:ACSHBPSv}. 

Identity \ref{id:II}+\ref{id:VI} ($\lambda=\mu$ and $\nu=1$):
\begin{align}
&\left.\frac{1}{\mu} r^{(d-1)\kappa} (1+G) f^{'\mu}\right|_0^\infty
+(d-1)\left(1-\frac{\kappa}{\mu}\right)
  \int_0^\infty\d{r}\; r^{(d-1)\kappa-1} (1+G) f^{'\mu} \non
&+\left(\frac{1}{2}-\frac{1}{\mu}\right)
  \int_0^\infty\d{r}\; r^{(d-1)\kappa} \frac{\p G}{\p f} f^{'\mu+1} 
+\left(1-\frac{1}{\mu}\right) 
  \int_0^\infty\d{r}\; r^{(d-1)\kappa} \frac{\p G}{\p r} f^{'\mu} \non
&-\int_0^\infty\d{r}\; r^{(d-1)\kappa} \frac{\p F}{\p f} f^{'\mu-1} 
-\int_0^\infty\d{r}\; r^{(d-1)\kappa} \frac{\p V}{\p f} f^{'\mu-1} = 0. 
\label{eq:idII+VI}
\end{align}
This simplifies the $(1+G)$ factors and removes the dependence on
$f''$. 

Identity \ref{id:I}+\ref{id:II}+\ref{id:VI} ($\lambda=\kappa=\mu$ and
$\nu=1$): 
\begin{align}
&\left.\frac{1}{\kappa}
  r^{(d-1)\kappa} (1+G) f^{'\kappa}\right|_0^\infty
+\left(\frac{1}{2}-\frac{1}{\kappa}\right)
  \int_0^\infty\d{r}\; r^{(d-1)\kappa} \frac{\p G}{\p f} f^{'\kappa+1} \non
&+\left(1-\frac{1}{\kappa}\right) 
  \int_0^\infty\d{r}\; r^{(d-1)\kappa} \frac{\p G}{\p r} f^{'\kappa} 
-\int_0^\infty\d{r}\; r^{(d-1)\kappa} \frac{\p F}{\p f} f^{'\kappa-1} \non
&-\int_0^\infty\d{r}\; r^{(d-1)\kappa} \frac{\p V}{\p f} f^{'\kappa-1} = 0. 
\label{eq:idI+II+VI}
\end{align}
In addition to the previous simplification, this one also eliminates
the second term and yields a one-parameter family of identities. 

Identity \ref{id:I}+\ref{id:VII} ($\lambda=\kappa$ and $\mu=1$):
\begin{align}
&\left.\frac{1}{\kappa}
  r^{(d-1)\kappa} (1+G)^\nu f'\right|_0^\infty 
+\left.\left(1-\frac{\nu}{\kappa}\right)
  r^{(d-1)\kappa} (1+G)^{\nu-1} \frac{\p G}{\p r}
  f\right|_0^\infty \non
&+\left(1-\frac{1}{\kappa}\right)
  \int_0^\infty\d{r}\; r^{(d-1)\kappa}(1+G)^\nu f'' 
+\left(\frac{1}{2}-\frac{\nu}{\kappa}\right)
  \int_0^\infty\d{r}\; r^{(d-1)\kappa}(1+G)^{\nu-1} \frac{\p G}{\p f}
  f^{'2} \non
&-(d-1)(\kappa-\nu)
  \int_0^\infty \d{r}\; r^{(d-1)\kappa-1} (1+G)^{\nu-1}
  \frac{\p G}{\p r} f \non
&-\left(1-\frac{\nu}{\kappa}\right)(\nu-1)
  \int_0^\infty \d{r}\; r^{(d-1)\kappa} (1+G)^{\nu-2}
  \frac{\p G}{\p r}
  \left(\frac{\p G}{\p f} f' + \frac{\p G}{\p r}\right) f \non
&-\left(1-\frac{\nu}{\kappa}\right) \int_0^\infty \d{r}\;
  r^{(d-1)\kappa} (1+G)^{\nu-1}
  \left(\frac{\p^2G}{\p f\p r}f' + \frac{\p^2 G}{\p r^2}\right) f 
-\int_0^\infty\d{r}\; r^{(d-1)\kappa} \frac{\p F}{\p f} (1+G)^{\nu-1} \non
&-\int_0^\infty\d{r}\; r^{(d-1)\kappa} (1+G)^{\nu-1} \frac{\p V}{\p f} = 0. 
\label{eq:idI+VII}
\end{align}
This identity possesses an extra boundary term for theories with
$G\neq 0$ when $\kappa\neq\nu$. This is the type-{\bf Ib} boundary
charge. 

Identity \ref{id:IV}+\ref{id:VII} ($\lambda=\nu$ and $\mu=1$):
\begin{align}
&\left.\frac{1}{\nu}
  r^{(d-1)\kappa} (1+G)^\nu f'\right|_0^\infty 
+\left.(d-1)\left(1-\frac{\kappa}{\nu}\right)
  r^{(d-1)\kappa-1} (1+G)^\nu f\right|_0^\infty \non
&-(d-1)\left(1-\frac{\kappa}{\nu}\right)\left[(d-1)\kappa-1\right]
  \int_0^\infty \d{r}\; r^{(d-1)\kappa-2} (1+G)^\nu f \non
&-(d-1)\left(1-\frac{\kappa}{\nu}\right)\nu
  \int_0^\infty \d{r}\; r^{(d-1)\kappa-1} (1+G)^{\nu-1}
  \frac{\p G}{\p f} f f' \non
&+\left(1-\frac{1}{\nu}\right)
  \int_0^\infty\d{r}\; r^{(d-1)\kappa}(1+G)^\nu f'' 
-\frac{1}{2} \int_0^\infty\d{r}\; r^{(d-1)\kappa}(1+G)^{\nu-1}
  \frac{\p G}{\p f} f^{'2} \non
&+(d-1)\left(\kappa-\nu\right)
  \int_0^\infty \d{r}\; r^{(d-1)\kappa-1} (1+G)^{\nu-1}
  \frac{\p G}{\p r} f 
-\int_0^\infty\d{r}\; r^{(d-1)\kappa} \frac{\p F}{\p f} (1+G)^{\nu-1} \non
&-\int_0^\infty\d{r}\; r^{(d-1)\kappa} (1+G)^{\nu-1} \frac{\p V}{\p f} = 0. 
\label{eq:idIV+VII}
\end{align}

Identity \ref{id:I}+\ref{id:IV}+\ref{id:VII} ($\lambda=\kappa=\nu$ and
$\mu=1$): 
\begin{align}
&\left.\frac{1}{\kappa}
  r^{(d-1)\kappa} (1+G)^\kappa f'\right|_0^\infty 
+\left(1-\frac{1}{\kappa}\right)
  \int_0^\infty\d{r}\; r^{(d-1)\kappa}(1+G)^\kappa f'' \non
&-\frac{1}{2} \int_0^\infty\d{r}\; r^{(d-1)\kappa}(1+G)^{\kappa-1}
  \frac{\p G}{\p f} f^{'2} 
-\int_0^\infty\d{r}\; r^{(d-1)\kappa} \frac{\p F}{\p f} (1+G)^{\kappa-1} \non
&-\int_0^\infty\d{r}\; r^{(d-1)\kappa} (1+G)^{\kappa-1} \frac{\p V}{\p f} = 0. 
\label{eq:idI+IV+VII}
\end{align}

Finally, another interesting combination is:\\
Identity \ref{id:VI}+\ref{id:VIII} ($\mu=2$ and $\nu=1$):
\begin{align}
&\left.\frac{1}{\lambda}
  r^{(d-1)\kappa} (1+G) f^{'2}\right|_0^\infty
\left.\mathop-r^{(d-1)\kappa} F\right|_0^\infty
\left.\mathop-r^{(d-1)\kappa} V\right|_0^\infty \non
&+(d-1)\left(1-\frac{\kappa}{\lambda}\right)
  \int_0^\infty\d{r}\; r^{(d-1)\kappa-1} (1+G) f^{'2} 
+\left(1-\frac{2}{\lambda}\right)
  \int_0^\infty\d{r}\; r^{(d-1)\kappa}(1+G) f' f'' \non
&+\left(\frac{1}{2}-\frac{1}{\lambda}\right)
  \int_0^\infty\d{r}\; r^{(d-1)\kappa} \frac{\p G}{\p f}
  f^{'3} 
+\left(1-\frac{1}{\lambda}\right)
  \int_0^\infty\d{r}\; r^{(d-1)\kappa} \frac{\p G}{\p r}
  f^{'2} 
+\int_0^\infty \d{r}\; r^{(d-1)\kappa} \frac{\p F}{\p r} \non
&+(d-1)\kappa \int_0^\infty \d{r}\; r^{(d-1)\kappa-1} F 
+(d-1)\kappa \int_0^\infty \d{r}\; r^{(d-1)\kappa-1} V 
=0.
\label{eq:idVI+VIII}
\end{align}
The above identity is particularly useful because it can be used to
calculate radial moments of the potential energy, which can be
interpreted as the size of the soliton.
In order to calculate the $n$-th radial moment of $V$, we set
$\kappa=\frac{d+n}{d-1}$ and get:
\begin{align}
&\left.\frac{1}{\lambda}
  r^{d+n} (1+G) f^{'2}\right|_0^\infty
\left.\mathop-r^{d+n} F\right|_0^\infty
\left.\mathop-r^{d+n} V\right|_0^\infty 
\non
&+\left(d-1-\frac{d+n}{\lambda}\right)
  \int_0^\infty\d{r}\; r^{d-1+n} (1+G) f^{'2} 
+\left(1-\frac{2}{\lambda}\right)
  \int_0^\infty\d{r}\; r^{d+n}(1+G) f' f'' \non
&+\left(\frac{1}{2}-\frac{1}{\lambda}\right)
  \int_0^\infty\d{r}\; r^{d+n} \frac{\p G}{\p f}
  f^{'3} 
+\left(1-\frac{1}{\lambda}\right)
  \int_0^\infty\d{r}\; r^{d+n} \frac{\p G}{\p r}
  f^{'2} 
+\int_0^\infty \d{r}\; r^{d+n} \frac{\p F}{\p r} \non
&+(d+n) \int_0^\infty \d{r}\; r^{d-1+n} F 
+(d+n) \int_0^\infty \d{r}\; r^{d-1+n} V 
=0.
\label{eq:idVI+VIII_moment}
\end{align}
A further simplification can be made by setting $\lambda=2$ yielding
identity \ref{id:II}+\ref{id:VI}+\ref{id:VIII} ($\lambda=\mu=2$ and
$\nu=1$) again with $\kappa=\frac{d+n}{d-1}$:
\begin{align}
&\left.\frac{1}{2}
  r^{d+n} (1+G) f^{'2}\right|_0^\infty
\left.\mathop-r^{d+n} F\right|_0^\infty
\left.\mathop-r^{d+n} V\right|_0^\infty 
\non
&+\left(\frac{d-n}{2}-1\right)
  \int_0^\infty\d{r}\; r^{d-1+n} (1+G) f^{'2} 
+\frac12 \int_0^\infty\d{r}\; r^{d+n} \frac{\p G}{\p r} f^{'2} 
+\int_0^\infty \d{r}\; r^{d+n} \frac{\p F}{\p r} \non
&+(d+n) \int_0^\infty \d{r}\; r^{d-1+n} F 
+(d+n) \int_0^\infty \d{r}\; r^{d-1+n} V 
=0.
\label{eq:idII+VI+VIII_moment}
\end{align}
This identity is particularly useful as it gives the $n$-th radial
moment of the potential ($V+F$) and it contains three boundary terms
-- of type {\bf 0}, type {\bf IIa} and type {\bf IIb}, respectively -- 
that may yield the topological number or other parameters of the
solution (see examples below). 

We note that in $d=2$ spatial dimensions, for the zeroth moment 
($n=0$), which is equivalent to the potential energy, the fourth term
(first term on the second line) vanishes; that case is also equal to
the combination \ref{id:I}+\ref{id:II}+\ref{id:VI}+\ref{id:VIII} of
the identities.

Up till now we have kept all identities completely general (of the
form given in Eq.~\eqref{eq:Lcankin}) so that they can be applied to a
vast number of theories possessing solitons in $d$ dimensions.
Although the capability of all the identities to yield physically or
mathematically interesting relations is not yet clear, we will show in
the examples in the next section that the various boundary terms
can yield either the topological number or other parameters
characterizing the soliton solutions, like the boundary conditions or, 
in principle, parameters like the shooting parameters.
In the last example above, we have shown that a particular combination
of the identity parameters ($\mu,\nu,\lambda$) yields a one-parameter
family of identities that generalize the famous Pohozaev
identity \cite{Pohozaev:1965}, although it is not guaranteed that it
relates the integrals to the topological number of the soliton.
In fact, we will see in the examples that it yields the winding number
for defect solitons, but not for texture solitons.

In general, the identities are meant to be applied with identity
parameters chosen such that all integrals converge.
In numerically difficult parts of parameter space, the identities may
be used to check the validity of the numerical solutions.
Another usage may be to reduce the number of integrals needed to
evaluate for instance the total energy. This may be useful for
obtaining a faster way of numerically calculate the (say) energy for a
large number of numerical solutions.
Finally, in some cases, solitons may have infinite energy but a finite
energy density. In such cases it may be difficult to check that the
numerical solutions are precise and in such case, the identities can
be used with a finite radial cut off $R$ and can then again be used to
check the numerics.

In the following sections we will demonstrate the identities with
some examples in various dimensions.

\section{Examples in two dimensions}\label{sec:example2}

\subsection{Abelian-Higgs BPS vortices}\label{sec:AHBPSv}

Let us consider the Abelian Higgs model at critical
coupling \cite{Jaffe:1980,Manton:2004}
\beq
\mathcal{L} = -\frac{1}{4e^2}F_{\alpha\beta}F^{\alpha\beta}
-(D_\alpha\phi)\overline{D^{\alpha}\phi}
-\frac{e^2}{2}(|\phi|^2-v^2)^2, 
\eeq
whose static energy density can by the Bogomol'nyi trick be written as
a sum of squares
\begin{equation}
\mathcal{E}^{\rm static} =
\frac{1}{2e^2}\left(F_{12} - e^2(|\phi|^2 - v^2)\right)^2
+ \left|D_1\phi + \i D_2\phi\right|^2
- v^2 F_{12}
- \i\epsilon^{ij}\frac{\p}{\p x^i}\big[\phi^\dag D_j\phi\big],
\end{equation}
where $D_i=\frac{\p}{\p x^i}+\i A_i$ is the gauge-covariant
derivative (note that the total-derivative term vanishes due to the
finite-energy condition $\lim_{r\to\infty}D_j\phi=0$),
$F_{12}=\frac{\p A_2}{\p x^1}-\frac{\p A_1}{\p x^2}$ is the field
strength, $A_i$ is the gauge potential and is a real two-vector,
$\phi\in\mathbb{C}$ is a complex-valued field, $e>0$ and $v>0$ are
real constants, and $i,j$ are summed over $1,2$.  
Setting the first two terms to zero yields the so-called BPS
equations and if we use the Ansatz
\beq
\phi = v h(r) \e^{\i N\theta}, \qquad
A_i = \epsilon_{ij} \frac{N x^j}{r^2} a(r),
\label{eq:vortex_ansatz}
\eeq
they read
\beq
h' = \frac{N}{r}(1-a)h, \qquad
\frac{N}{r} a' = e^2 v^2 (1 - h^2),
\label{eq:ANO_BPS}
\eeq
where we have used the standard polar coordinates
$x^1+\i x^2=r\e^{\i\theta}$. 
Combining the above two BPS equations, we get the master equation
\beq
f'' + \frac{1}{r}f' + m^2(1 - \e^f) = 0,
\label{eq:Taubes}
\eeq
which is the famous Taubes equation \cite{Taubes:1979tm} and we have
defined 
\beq
f \equiv 2\log h, \qquad
m \equiv \sqrt{2} e v.
\eeq
The Taubes equation has an auxiliary Lagrangian density
\beq
\mathcal{L} =
-\frac{1}{2}f^{'2}
-m^2(\e^f - 1 - f),
\eeq
whose equation of motion is exactly the Taubes
equation \eqref{eq:Taubes}.
The solutions of this equation are called BPS vortices and are prime
examples of topological defects. 

Comparing to the Lagrangian density \eqref{eq:Lcankin}, we can
identify
\beq
F = G = 0, \qquad
V = m^2(\e^f - 1 - f).
\eeq
Finally, we need to know the behaviors of $f$ at large and small radii
in order to evaluate the total derivative terms in the identity. 
The solutions obey the boundary conditions $h(0)=0$ and
$h(\infty)=1$. The behavior at small $r$ for $h$ is
$h=Ar^N+\mathcal{O}(r^{N+2})$ and thus $f=2N\log r + 2\log A$.
At large $r$, we can linearize Eq.~\eqref{eq:Taubes} (because $f$
needs to be close to its vacuum value, $0$, at large radii) to get the 
solution $f=-B K_0(m r)$, which behaves like
$f=-B\sqrt{\frac{\pi}{2 m r}} \e^{-m r}$ at large $r$.  

Using the identity \ref{id:V} \eqref{eq:idV}, we get
\begin{align}
\frac{1}{\lambda}(2N)^\mu \delta^{\kappa\mu}
&= m^2 \int_0^\infty \d{r}\; r^\kappa (1 - \e^f) f^{'\mu-1}
-\left(\frac{\kappa}{\lambda}-1\right) \int_0^\infty \d{r}\;
  r^{\kappa-1} f^{'\mu} \non
&\phantom{=\ }
+\left(1-\frac{\mu}{\lambda}\right)\int_0^\infty \d{r}\; r^\kappa
  f^{'\mu-1} f'',
\label{eq:ANO_BPS_idV}
\end{align}
valid for $\kappa\geq\mu>1$, $\delta$ is the Kronecker-delta,
and $\lambda$ is a free real parameter.
One can show that all solutions satisfy $f\leq 0$ and $f'\geq 0$ and
hence the first two integrals on the right-hand side are positive
definite. 

This is a prime example of an identity yielding a type-{\bf 0} boundary
charge, which in this case is proportional to $N^\mu$ when
$\kappa=\mu$.  

A natural choice is to set $\lambda=\mu$ to get
identity \ref{id:II}+\ref{id:V}:
\begin{align}
\frac{1}{\mu}(2N)^\mu \delta^{\kappa\mu}
&= m^2 \int_0^\infty \d{r}\; r^\kappa (1 - \e^f) f^{'\mu-1}
-\left(\frac{\kappa}{\mu}-1\right) \int_0^\infty \d{r}\;
  r^{\kappa-1} f^{'\mu}.
\label{eq:ANO_BPS_idII+V}
\end{align}
A special case is $\kappa=\mu$
(i.e.~identity \ref{id:I}+\ref{id:II}+\ref{id:V}), which gives 
infinitely many integrals simply related to the topological winding
number $N$: 
\beq
N^\kappa = 
\frac{\kappa m^2}{2^\kappa} \int_0^\infty \d{r}\;
  r^\kappa (1 - \e^f) f^{'\kappa-1},
\label{eq:ANO_BPS_master_id}
\eeq
for $\kappa\geq 1$.
Consider in particular the cases $\kappa=1,2$
(i.e.~identities \ref{id:I}+\ref{id:II}+\ref{id:V}+\ref{id:VII}
and \ref{id:I}+\ref{id:II}+\ref{id:V}+\ref{id:VIII}, respectively),
for which we can write
\begin{align}
N &= \frac{m^2}{2} \int_0^\infty \d{r}\; r (1-\e^f),
  \label{eq:BPSvortexN}\\
N^2 &= \frac{m^2}{2} \int_0^\infty \d{r}\; r^2 (1-\e^f) f'
     = -m^2 \int_0^\infty \d{r}\; r(1-\e^f+f),
  \label{eq:BPSvortexNsq}
\end{align}
Comparing the above two equations, we can determine the area integral
of $f$ as
\beq
N(N+2) = - m^2\int_0^\infty \d{r}\; r f > 0.
\label{eq:BPSvortexNsqplus2N}
\eeq

Two more special cases arise, because we can integrate the two
integrals by parts in Eq.~\eqref{eq:ANO_BPS_idII+V} for $\mu=2$
and $\mu=1$, respectively.
Let us write them out, explicitly, as
identity \ref{id:II}+\ref{id:V}+\ref{id:VIII}: 
\beq
N^2\delta^{\kappa 2}
= \frac{\kappa m^2}{2}\int_0^\infty \d{r}\; r^{\kappa-1} (\e^f - 1 - f) 
-\frac{1}{2}\left(\frac{\kappa}{2} - 1\right) \int_0^\infty \d{r}\;
  r^{\kappa-1} f^{'2}, \quad
\kappa\geq 2,
\eeq
and identity \ref{id:II}+\ref{id:V}+\ref{id:VII}:
\beq
N\delta^{\kappa 1}
= \frac{m^2}{2}\int_0^\infty \d{r}\; r^\kappa (1 - \e^f)
+ \frac{1}{2}(\kappa-1)^2 \int_0^\infty \d{r}\; r^{\kappa-2} f, \quad
\kappa\geq 1.
\eeq
The above expression gives the $n$-th radial moment of $f$ in terms of
the $(n+2)$-th moment of $\frac{\p V}{\p f}$.
The case $\kappa=1$ reduces to
Eq.~\eqref{eq:BPSvortexN}.

For some of the identities, one may consider forming a geometric
series and summing it up, yielding new functional forms.\footnote{We
thank Ken Konishi for this suggestion.}
As an example, we can multiply Eq.~\eqref{eq:ANO_BPS_master_id} by
$\beta^\kappa$ and sum over $\kappa$:
\begin{align}
\sum_{\kappa=1}^\infty (\beta N)^\kappa
= m^2 \int_0^\infty \d{r}\; \frac{(1-\e^f)}{f'}
  \sum_{\kappa=1}^\infty
  \frac{\kappa \left(\beta r f'\right)^\kappa}{2^\kappa},
\end{align}
and so we arrive at
\beq
\frac{\beta N}{1 - \beta N}
= 2\beta m^2 \int_0^\infty \d{r}\; \frac{r(1-\e^f)}{(\beta r f' -
2)^2},
\label{eq:BPSvortexresummed}
\eeq
where $0<\beta<\frac{1}{N}$ is a small real parameter.
Note that we assume $\beta r f'<2$, which is always possible for small
enough values of $\beta$ as $f'$ asymptotically is exponentially
suppressed.

\subsection{Abelian Chern-Simons-Higgs BPS vortices}\label{sec:ACSHBPSv}

Let us consider the Abelian Chern-Simons-Higgs model at critical
coupling (see e.g.~\cite{Dunne:1995,Yang:2001,Tarantello:2008}), 
\beq
\mathcal{L} = -\frac{k}{4}\epsilon^{\alpha\beta\gamma}
  A_\alpha F_{\beta\gamma}
-(D_\alpha\phi)\overline{D^\alpha\phi}
- \frac{1}{k^2}|\phi|^2(|\phi|^2-v^2)^2,
\eeq
whose energy density
\beq
\mathcal{E} = |D_0\phi|^2 + |D_i\phi|^2
+ \frac{1}{k^2}|\phi|^2(|\phi|^2 - v^2),
\eeq
can be written by Bogomol'nyi completion as
\beq
\mathcal{E} =
\left|D_0\phi - \frac{\i}{k}\phi(|\phi|^2 - v^2)\right|^2
+ |D_1\phi + \i D_2\phi|^2
- v^2 F_{12}
- \i\epsilon^{ij}\frac{\p}{\p x^i}\big[\phi^\dag D_j\phi\big],
\label{eq:AbelianCS_Bogmolnyi}
\eeq
which holds due to the Gauss' law
\beq
k F_{12} = -\i\phi^\dag D_0 \phi + \i\phi (D_0 \phi)^\dag,
\eeq
and $k$ and $v$ are real nonzero parameters (which we take to be
positive here). $\phi\in\mathbb{C}$ is still a complex-valued field,
but the gauge field $A_\alpha$ now is a real three-vector; this is
because -- as we shall see shortly -- Gauss' law relates a magnetic
flux with an electric charge for nonzero $k>0$, which in turn requires
a nontrivial $A_0$ component of the gauge field.

Solving the above equation for $A_0$,
\beq
A_0 = \frac{k F_{12}}{2|\phi|^2}
+ \frac{\i\phi^\dag\frac{\p\phi}{\p x^0}
  - \i\phi\frac{\p\phi^\dag}{\p x^0}}{2|\phi|^2},
\eeq
and substituting into the BPS equations (obtained by setting the first
two terms in Eq.~\eqref{eq:AbelianCS_Bogmolnyi} to zero), we get the
static equations 
\beq
F_{12} = \frac{2}{k^2}|\phi|^2(|\phi|^2 - v^2), \qquad
D_1\phi + \i D_2\phi = 0.
\eeq
Using the Ansatz \eqref{eq:vortex_ansatz}, the above BPS equations
read
\beq
h' = \frac{N}{r}(1-a)h, \qquad
\frac{N}{r} a' = \frac{2v^4}{k^2} (1 - h^2) h^2,
\label{eq:AbelianCS_BPS}
\eeq
which can be combined into the master equation
\beq
f'' + \frac{1}{r}f' + m^2 (1 - \e^f) \e^f = 0,
\label{eq:ACS_master}
\eeq
where we have defined
\beq
f\equiv 2\log h, \qquad
m \equiv \frac{2v^2}{k}.
\label{eq:CSmass}
\eeq
The master equation \eqref{eq:ACS_master} has the following auxiliary
Lagrangian density 
\beq
\mathcal{L} = -\frac{1}{2}f^{'2}
- \frac{m^2}{2} (1 - \e^f)^2,
\eeq
whose equation of motion is exactly the master equation.
The solutions of this master equation are called BPS Chern-Simons
vortices and also topological defects. 
Comparing to the Lagrangian \eqref{eq:Lcankin}, we identify
\beq
F = G = 0, \qquad
V = \frac{m^2}{2} (1 - \e^f)^2.
\eeq
Finally, we need the behaviors of $f$ at small and large radii.
It is easy to see that they are exactly the same in this case as in
the previous section and hence $f=2N\log r+2\log A$ at small $r$ and 
$f=-B K_0(m r)$ at large $r$ (with the mass defined in
Eq.~\eqref{eq:CSmass}). 

We are now ready to apply the identity. Let us start with the most
general one, i.e.~identity \ref{id:V} \eqref{eq:idV} (as $G$
vanishes):
\begin{align}
\frac{1}{\lambda} (2N)^\mu \delta^{\kappa\mu}
&= m^2\int_0^\infty \d{r}\; r^\kappa (1 - \e^f) \e^f f^{'\mu-1}
-\left(\frac{\kappa}{\lambda} - 1\right) \int_0^\infty \d{r}\;
  r^{\kappa-1} f^{'\mu} \non
&\phantom{=\ }
+\left(1 - \frac{\mu}{\lambda}\right) \int_0^\infty \d{r}\; r^\kappa
  f^{'\mu-1} f'',
\label{eq:AbelianCS_BPS_idV}
\end{align}
valid for $\kappa\geq\mu>1$. 
Again a natural choice is to set $\lambda=\mu$ to get
identity \ref{id:II}+\ref{id:V}: 
\beq
\frac{1}{\mu} (2N)^\mu \delta^{\kappa\mu}
= m^2\int_0^\infty \d{r}\; r^\kappa (1 - \e^f) \e^f f^{'\mu-1}
- \left(\frac{\kappa}{\mu} - 1\right) \int_0^\infty \d{r}\;
  r^{\kappa-1} f^{'\mu}.
\label{eq:AbelianCS_BPS_idII+V}
\eeq
This is again an example of an identity yielding a type-{\bf 0} boundary 
charge, which is proportional to $N^\mu$ when $\kappa=\mu$.

As in the case of the standard Abelian BPS vortices, a special case is
$\kappa=\mu$ (i.e.~identity \ref{id:I}+\ref{id:II}+\ref{id:V}), which
gives infinitely many integrals simply related to the winding number
$N$:  
\beq
N^\kappa = \frac{\kappa m^2}{2^\kappa}
\int_0^\infty \d{r}\; r^\kappa (1 - \e^f) \e^f f^{'\kappa-1},
\label{eq:CS_BPS_master_id}
\eeq
for $\kappa\geq 1$. 
Let us write out a few examples of $\kappa=1,2$, explicitly
\begin{align}
N &= \frac{m^2}{2}\int_0^\infty \d{r}\; r (1-\e^f) \e^f,
\label{eq:BPSAbelianCSvortexN}\\
N^2 &= \frac{m^2}{2}\int_0^\infty \d{r}\; r^2 (1-\e^f)\e^f f'
     = \frac{m^2}{2} \int_0^\infty \d{r}\; r (1-\e^f)^2.
\label{eq:BPSAbelianCSvortexNsq}
\end{align}
These two relations can also be found in Ref.~\cite{Tarantello:2008}.
The first one is well known as it is the winding number where the BPS
equation has been used.
Comparing the above two equations, we get
\beq
N(N+1) = \frac{m^2}{2}\int_0^\infty \d{r}\; r(1- \e^f),
\label{eq:BPSAbelianCSvortexNsqplusN}
\eeq
and
\beq
N(N+2) = \frac{m^2}{2}\int_0^\infty \d{r}\; r(1-\e^{2f}).
\label{eq:BPSAbelianCSvortexNsqplus2N}
\eeq

Again, two more special cases arise because we can integrate the two
integrals in Eq.~\eqref{eq:AbelianCS_BPS_idII+V} by parts for
$\mu=2$ and $\mu=1$, respectively.
Writing them out explicitly, we get
identity \ref{id:II}+\ref{id:V}+\ref{id:VIII}: 
\beq
N^2\delta^{\kappa 2}
= \frac{\kappa m^2}{4} \int_0^\infty \d{r}\; r^{\kappa-1} (1 - \e^f)^2
-\frac{1}{2}\left(\frac{\kappa}{2} - 1\right)
  \int_0^\infty \d{r}\; r^{\kappa-1} f^{'2}, \qquad
\kappa\geq 2,
\eeq
and identity \ref{id:II}+\ref{id:V}+\ref{id:VII}:
\beq
N \delta^{\kappa 1}
= \frac{m^2}{2} \int_0^\infty \d{r}\; r^\kappa (1 - \e^f) \e^f
+ \frac{1}{2} (\kappa-1)^2 \int_0^\infty \d{r}\; r^{\kappa-2} f, \qquad
\kappa\geq 1.
\eeq
The latter expression again gives the $n$-th radial moment of $f$ in
terms of the $(n+2)$-th moment of $\frac{\p V}{\p f}$. As a
consistency check, we can see that $\kappa=1$ reduces the latter to 
Eq.~\eqref{eq:BPSAbelianCSvortexN} and $\kappa=2$ reduces the first to
Eq.~\eqref{eq:BPSAbelianCSvortexNsq}.

\subsection{Global Abelian vortices}\label{sec:globalvtx}

Let us now consider the global (i.e.~ungauged) Abelian
vortex \cite{Vilenkin:1994,Manton:2004} described by the Lagrangian 
\beq
\mathcal{L} =
- (\p_\alpha\phi)\overline{\p^\alpha\phi}
- \frac{\xi}{4}(|\phi|^2 - v^2)^2,
\eeq
where $\phi\in\mathbb{C}$ is a complex-valued field,
$\p_\alpha\phi\equiv\frac{\p\phi}{\p x^\alpha}$ and $\xi>0$, $v>0$
are two positive real parameters.  
Inserting the Ansatz for a vortex with winding number
$N\in\mathbb{Z}_{>0}$,
\beq
\phi = v f(r) \e^{\i N\theta},
\eeq
and using rescaled coordinates
$x^1 + \i x^2=\frac{\sqrt{2}r}{v}\e^{\i\theta}$, 
we get the reduced Lagrangian density
\beq
\frac{1}{v^4}\mathcal{L} =
- \frac{1}{2}f^{'2}
- \frac{N^2}{2r^2}f^2
- \frac{\xi}{4}(f^2 - 1)^2,
\eeq
which is the simplest example in two dimensions.
The prime denotes derivative with respect to the radial coordinate as
usual $f'=\frac{\d{f}}{\d{r}}$. 
Unfortunately, as well-known, the global vortex has infinite energy
(logarithmically divergent with the cut-off) due to the second term in
the above Lagrangian density. 
The coincident (axially symmetric) $N$-vortex with $N>1$ is also
unstable; here we will leave $N$ as a free parameter of the system,
but the reader should bear in mind that the multivortex ($N>1$) will
eventually decay into $N$ spatially separated 1-vortices.

Comparing to the Lagrangian density \eqref{eq:Lcankin}, we can
identify
\beq
F = \frac{N^2}{2r^2} f^2, \qquad
G = 0, \qquad
V = \frac{\xi}{4}(f^2 - 1)^2.
\eeq

The equation of motion is
\beq
f'' + \frac{1}{r}f'
- \frac{N^2}{r^2}f
- \xi(f^2-1)f = 0.
\eeq
The solutions of the above equation are called global vortices and
are topological defects. 
The asymptotic behavior of $f$ for $r\to\infty$ can be found by
linearizing the above equation around $f=1-\delta$, which gives the
solution
\beq
f = 1 - B K_N(\sqrt{2\xi}r),
\eeq
which at asymptotically large radii is well approximated by
\beq
f = 1 - \sqrt{\frac{\pi}{2\sqrt{2\xi}r}} B \e^{-\sqrt{2\xi} r},
\eeq
and hence $f'$ goes to zero as
$f'\simeq\sqrt{\frac{\pi\sqrt{2\xi}}{2r}}B\e^{-\sqrt{2\xi}r}$.
At small $r$, the behavior is
\beq
f = A r^N + \mathcal{O}(r^{N+2}).
\eeq

We are now ready to apply the identity. Since $G=0$, we start with the
most general case, which is given by identity \ref{id:V}:
\begin{align}
\left(\frac{\kappa}{\lambda}-1\right) \int_0^\infty \d{r}\; r^{\kappa-1}
  f^{'\mu}
+\left(\frac{\mu}{\lambda}-1\right)\int_0^\infty \d{r}\; r^\kappa
  f^{'\mu-1} f''
+N^2 \int_0^\infty \d{r}\; r^{\kappa-2} f f^{'\mu-1} \non
-\xi \int_0^\infty \d{r}\; r^\kappa (1-f^2)f f^{'\mu-1} = 0,
\end{align}
which is valid for $\kappa+\mu(N-1)>0$ and $\mu>1$. 
Note that there is no type-{\bf 0} boundary charge in this case.

A natural choice is to set $\lambda=\mu$, which is
identity \ref{id:II}+\ref{id:V}:
\begin{equation}
\left(\frac{\kappa}{\mu} - 1\right) \int_0^\infty \d{r}\; r^{\kappa-1}
  f^{'\mu}
+ N^2\int_0^\infty \d{r}\; r^{\kappa-2} f f^{'\mu-1}
- \xi\int_0^\infty \d{r}\; r^\kappa (1 - f^2) f f^{'\mu-1} = 0.
\label{eq:global_vortex_identity1_I+IV}
\end{equation}

A special case is again $\kappa=\mu$, which is
identity \ref{id:I}+\ref{id:II}+\ref{id:V}: 
\beq
N^2\int_0^\infty \d{r}\; r^{\kappa-2} f f^{'\kappa-1}
=\xi\int_0^\infty \d{r}\; r^{\kappa} (1-f^2)f f^{'\kappa-1},
\eeq
which is valid for $\kappa>1$.
In particular, for $\kappa=2$, we can integrate by parts to get
identity \ref{id:I}+\ref{id:II}+\ref{id:V}+\ref{id:VIII}:
\beq
N^2 = \xi\int_0^\infty \d{r}\; r(1-f^2)^2,
\label{eq:globalvtx_Derrick-Pohozaev}
\eeq
which is the well-known Pohozaev
identity \cite{Pohozaev:1965}, relating the potential energy to the 
winding number of the vortex solution.
In our language, the latter is a type-{\bf IIa} boundary charge.
This identity is especially useful, since the total energy is
infinite, hence this finite quantity can be used to check the accuracy
of the solutions.

Similarly to the other models, also here two more special cases arise
where we can integrate the first and the last two integrals by parts
in Eq.~\eqref{eq:global_vortex_identity1_I+IV} for $\mu=1$ and
$\mu=2$, respectively.
Starting with the $\mu=2$ case, writing it out explicitly we get
identity \ref{id:II}+\ref{id:V}+\ref{id:VIII} ($\mu=2$): 
\begin{align}
-\int_0^\infty \frac{\d r}{r} f^{'2}
+ N^2\int_0^\infty \frac{\d r}{r^3} f^2 &= \frac{\xi}{4}, \qquad
N>1,\label{eq:global_vortex_identity1_IV+VIa}\\
\left(1 - \frac{\kappa}{2}\right) \int_0^\infty \d{r}\; r^{\kappa-1}
  f^{'2}
-\left(1 - \frac{\kappa}{2}\right)N^2\int_0^\infty \d{r}\; r^{\kappa-3} f^2 &\non
\mathop+\frac{\kappa\xi}{4}\int_0^\infty \d{r}\; r^{\kappa-1}
  (f^2-1)^2 &= 0, \qquad 0<\kappa<2,\\
\left(\frac{\kappa}{2} - 1\right) \int_0^\infty \d{r}\; r^{\kappa-1}
  f^{'2} 
- \left(\frac{\kappa}{2} - 1\right) N^2\int_0^\infty \d{r}\; r^{\kappa-3} (f^2 - 1) &\non
\mathop- \frac{\kappa\xi}{4} \int_0^\infty \d{r}\; r^{\kappa-1}
  (f^2 - 1)^2 &= -\frac{N^2}{2}\delta^{\kappa 2}, \qquad \kappa\geq 2.
\label{eq:global_vortex_identity1_IV+VIc}
\end{align}
Two boundary charges of type {\bf IIb} and type {\bf IIa} are recovered in 
Eq.~\eqref{eq:global_vortex_identity1_IV+VIa} and
Eq.~\eqref{eq:global_vortex_identity1_IV+VIc}, respectively.

Eq.~\eqref{eq:global_vortex_identity1_IV+VIc} gives the $n$-th radial
moment (with $n=\kappa-2$) of the terms in the Lagrangian density;
more precisely, let us define
\beq
\frac{\widetilde{\mathcal{L}}}{v^4} \equiv
+ \frac{1}{2}f^{'2}
- \frac{N^2}{2r^2}f^2
- \frac{\xi}{4}(f^2 - 1)^2,
\eeq
in terms of which we can write
\begin{equation}
\frac{\kappa-2}{v^4}\int_0^R \d{r}\; r^{\kappa-1} \widetilde{\mathcal{L}}
= -\frac{N^2}{2}\delta^{\kappa2}
-\frac{N^2}{2}(\kappa-2) \int_0^R \d{r}\; r^{\kappa-3}
+ \frac{\xi}{2}\int_0^R \d{r}\; r^{\kappa-1} (f^2-1)^2,
\end{equation}
for $R\to\infty$.
The special case $\kappa=2$ is, of course, still the Pohozaev
identity. 
Obviously, the above equation diverges on both sides (unlike all terms
in
Eqs.~(\ref{eq:global_vortex_identity1_IV+VIa}-\ref{eq:global_vortex_identity1_IV+VIc}))
since $\widetilde{\mathcal{L}}$ is just as divergent as
$\mathcal{L}$. 
Nevertheless, for a finite cut-off radius $R$, we can calculate each
side and treat $R$ as an infrared regulator; in this case the above
equation can be used to test numerical solutions for large enough $R$  
(that is, in the regime where $(1-f)\ll 1$).
We note that $\widetilde{\mathcal{L}}$ is not the energy (minus the 
Lagrangian density), but it is the Legendre transform of the
Lagrangian in the radial direction; it does not have any known
interpretation in physics though (recall the Hamiltonian is the
Legendre transform of the Lagrangian in the temporal direction). 
The divergence on the left-hand side comes from the second term in
$\widetilde{\mathcal{L}}$ while on the right-hand side it also comes
from the second term. 
The advantage of the original form of
Eqs.~(\ref{eq:global_vortex_identity1_IV+VIa}-\ref{eq:global_vortex_identity1_IV+VIc})
is that every term is convergent.

We consider now the other case, i.e.~$\mu=1$, for which we get
identity \ref{id:II}+\ref{id:V}+\ref{id:VI}: 
\begin{align}
&\left(N^2 - (\kappa-1)^2\right) \int_0^\infty \d{r}\; r^{\kappa-2} f
= \xi \int_0^\infty \d{r}\; r^\kappa (1 - f^2)f, \qquad
-N+1<\kappa<1.
\label{eq:globalvtx_idII+V+VI}
\end{align}
This equation is an example where we utilize a negative range of
$\kappa$ by having a large enough winding number $N>1$.
For the physically stable case $N=1$, $\kappa$ remains restricted to
the positive range $0<\kappa<1$, of course. 

\subsection{Baby Skyrmions}\label{sec:babyskyrmions}

Let us next consider the baby-Skyrme model which is a 2-dimensional
Skyrme model using just an O(3) vector
field \cite{Leese:1989gi,Piette:1994jt,Piette:1994ug}, whereas the full Skyrme
model (see Sec.~\ref{sec:Skyrmions}) is a full 3-dimensional model
with an O(4) vector field. 
The Lagrangian density reads
\beq
\mathcal{L} =
-\frac{1}{2}\p_\alpha\mathbf{n}\cdot\p^\alpha\mathbf{n}
-\frac{c_4}{4}(\p_\alpha\mathbf{n}\times\p_\beta\mathbf{n})\cdot
  (\p^\alpha\mathbf{n}\times\p^\beta\mathbf{n}) - V[n^3],
\label{eq:LbabySkyrmion}
\eeq
where $\p_\alpha\mathbf{n}\equiv\frac{\p\mathbf{n}}{\p x^\alpha}$,
$\mathbf{n}\cdot\mathbf{n}=\sum_{a=1}^3 n^a n^a=1$, $c_4>0$ is a
constant and finally, the potential is only a functional of the third
component of the vector field $\mathbf{n}$.
Using the Ansatz
\beq
\mathbf{n} =
(\sin f(r)\cos(N\theta),
\sin f(r)\sin(N\theta),
\cos f(r)),
\eeq
the reduced Lagrangian density simplifies to
\beq
\mathcal{L} =
-\frac{1}{2}f^{'2}
-\frac{N^2}{2r^2}\sin^2 f
-\frac{c_4N^2}{2r^2}\sin^2(f) f^{'2}
-V[\cos f],
\label{eq:reducedLbabySkyrmions}
\eeq
where $N$ is the topological degree and also the number of
baby Skyrmions. Baby Skyrmions are an example of topological
textures. 
We should again warn the reader, that the baby Skyrmions for $N>1$
energetically prefer to form a chain or other shapes, instead of an
axially symmetric
configuration \cite{Foster:2009vk,Hen:2007in,Weidig:1998ii} (although
the details depend on the potential, of course).
It will, nevertheless, be instructive to leave $N$ as a free parameter
in the following, to get a feel for where the topological degree plays
a role. 

The equation of motion is
\begin{align}
&f'' + \frac{1}{r}f'
+\frac{c_4N^2}{r^2}\sin^2(f)f''
-\frac{c_4N^2}{r^3}\sin^2(f)f'
+\frac{c_4N^2}{2r^2}\sin(2f)f^{'2}
-\frac{N^2}{2r^2}\sin 2f \non
& + \sin(f) V'[\cos f] = 0,
\end{align}
and the boundary conditions for the baby Skyrmions are $f(0)=\pi$ and 
$f(\infty)=0$. 

Comparing the Lagrangian density \eqref{eq:reducedLbabySkyrmions} to
Eq.~\eqref{eq:Lcankin}, we can read off
\beq
F = \frac{N^2}{2r^2} \sin^2 f, \qquad
G = \frac{c_4 N^2}{r^2} \sin^2 f, \qquad
V = V.
\eeq
This is the first example with a non-vanishing $G$ functional.

There are many interesting potentials utilized in the literature;
therefore we will keep $V$ general in the following. Among a few
possibilities is the standard mass term
\beq
V = m^2 (1 - \cos f),
\label{eq:VbabySk_standard}
\eeq
or the modified mass term \cite{Kudryavtsev:1997nw,Weidig:1998ii} 
\beq
V = \frac{1}{2} m^2 (1 - \cos^2 f) = \frac{1}{2} m^2 \sin^2 f,
\label{eq:VbabySk_modified}
\eeq
or the ``holomorphic'' mass term \cite{Leese:1989gi,Sutcliffe:1991aua}
\beq
V = \frac{1}{4}m^2(1 - \cos f)^4,
\label{eq:VbabySk_holomorphic}
\eeq
or a linear combination of the standard mass and holomorphic mass
term
\beq
V = m^2 \alpha (1 - \cos f) + m^2 (1-\alpha)(1 - \cos f)^4,
\label{eq:VbabySk_aloof}
\eeq
dubbed the aloof mass term in Ref.~\cite{Salmi:2014hsa} due to its
properties that entails short-distance repulsion and long-distance
attraction among spatially separated baby Skyrmions.

We should mention that due to the kinetic term being (classically)
conformal (scale invariant), stability of the model requires a
potential and we will require the potential to have a global vacuum
(not necessarily the only one) at $f=0$. 

Finally, we need the behavior of the field $f$ at small and large
radii.
At small $r$, one can determine
\beq
f = \pi - A r^N + \mathcal{O}(r^{N+2}),
\eeq
while at large $r$ it depends crucially on the potential.
For the massive case, corresponding to
Eqs.~\eqref{eq:VbabySk_standard}, \eqref{eq:VbabySk_modified}
and \eqref{eq:VbabySk_aloof} with $\alpha>0$, we get by linearizing
the equation of motion \eqref{eq:reducedLbabySkyrmions}:
\beq
f''
+\frac{1}{r} f'
-\frac{N^2}{r^2} f
+f \left(\left.V'[\cos f]\right|_{f=0}\right) = 0,
\eeq
the solution
\beq
f \simeq B K_N(m r),
\eeq
but in the case of \eqref{eq:VbabySk_standard}, the mass is modified
as $m\to\sqrt{\alpha}m$.
In the massless case, which corresponds -- in the examples we outlined 
above -- to Eq.~\eqref{eq:VbabySk_holomorphic}, the fall off becomes 
polynomial
\beq
f \simeq \frac{B}{r^N}.
\eeq
Therefore, we need to analyze the massive and the massless
cases separately in the following.
That means the valid range of the powers $\mu$, $\nu$ and $\kappa$
depends on whether the potential is of massive or massless type.

Let us apply the identity \eqref{eq:id}, obtaining
\begin{align}
\left(1 - \frac{\kappa}{\lambda}\right)
  \int_0^\infty \d{r}\; r^{\kappa-1} (1+G)^\nu f^{'\mu} 
+\left(1 - \frac{\mu}{\lambda}\right) \int_0^\infty \d{r}\;
  r^\kappa (1+G)^\nu f^{'\mu-1} f'' \non 
\mathop+c_4 N^2\left(\frac12 - \frac{\nu}{\lambda}\right) 
  \int_0^\infty \d{r}\; r^{\kappa-2} (1+G)^{\nu-1} \sin(2f)
  f^{'\mu+1} \non
\mathop-2c_4N^2\left(1 - \frac{\nu}{\lambda}\right) \int_0^\infty \d{r}\;
  r^{\kappa-3} (1+G)^{\nu-1} \sin^2f f^{'\mu} \non
\mathop-\frac{N^2}{2}\int_0^\infty \d{r}\; r^{\kappa-2}
  (1+G)^{\nu-1} \sin(2f) f^{'\mu-1} 
-\int_0^\infty \d{r}\; r^\kappa (1+G)^{\nu-1}
  \frac{\p V}{\p f} f^{'\mu-1} = 0,
\label{eq:babySk_id}
\end{align}
generically valid for $\kappa>\mu(1-N)$ and $\mu>0$.

As the potential is unspecified at this point, it will also provide a
constraint on $\kappa$ for the massless case.
Let us for concreteness consider the potentials mentioned above.
For the potentials
\eqref{eq:VbabySk_standard}, \eqref{eq:VbabySk_modified}
and \eqref{eq:VbabySk_aloof} the field is massive and there is thus no
upper bound on $\kappa$ at all.
For the ``massless'' potential or so-called holomorphic
potential \eqref{eq:VbabySk_holomorphic}, the field is massless and
the upper limit on $\kappa$ coming from the potential is
$\kappa<7N-1$; that bound is weaker than that coming from the other
terms in Eq.~\eqref{eq:babySk_nu1_mu1} and so the upper
bound on $\kappa$ remains $\kappa<N+1$ in this case.

As usual a natural choice is to eliminate $f''$ by setting
$\lambda=\mu$ obtaining identity \ref{id:II}:
\begin{align}
\left(1 - \frac{\kappa}{\mu}\right)
  \int_0^\infty \d{r}\; r^{\kappa-1} (1+G)^\nu f^{'\mu} 
+c_4 N^2\left(\frac12 - \frac{\nu}{\mu}\right) 
  \int_0^\infty \d{r}\; r^{\kappa-2} (1+G)^{\nu-1} \sin(2f)
  f^{'\mu+1} \non
\mathop-2c_4N^2\left(1 - \frac{\nu}{\mu}\right) \int_0^\infty \d{r}\;
  r^{\kappa-3} (1+G)^{\nu-1} \sin^2f f^{'\mu} \non
\mathop-\frac{N^2}{2}\int_0^\infty \d{r}\; r^{\kappa-2}
  (1+G)^{\nu-1} \sin(2f) f^{'\mu-1} 
-\int_0^\infty \d{r}\; r^\kappa (1+G)^{\nu-1}
  \frac{\p V}{\p f} f^{'\mu-1} = 0.
\label{eq:babySk_idII}
\end{align}

Let us consider the simplification provided by
identity \ref{id:II}+\ref{id:VI} (i.e.~setting $\nu=1$):
\begin{align}
\left(1 - \frac{\kappa}{\mu}\right)
  \int_0^\infty \d{r}\; r^{\kappa-1} f^{'\mu} 
\mathop+c_4 N^2\left(\frac12 - \frac{1}{\mu}\right) 
  \int_0^\infty \d{r}\; r^{\kappa-2} \sin(2f) f^{'\mu+1} \non
\mathop-c_4N^2\left(1 + \frac{\kappa}{\mu} - \frac{2}{\mu}\right)
  \int_0^\infty \d{r}\; r^{\kappa-3} \sin^2(f) f^{'\mu} 
-\frac{N^2}{2}\int_0^\infty \d{r}\; r^{\kappa-2} \sin(2f) f^{'\mu-1} \non
-\int_0^\infty \d{r}\; r^\kappa \frac{\p V}{\p f} f^{'\mu-1} = 0.
\label{eq:babySk_nu1}
\end{align}
We can simplify this identity further by setting $\mu=1$ or $\mu=2$.
Let us start with the $\mu=1$ case, which is
identity \ref{id:II}+\ref{id:VI}+\ref{id:VII}: 
\begin{align}
(1 - \kappa)^2\int_0^\infty \d{r}\; r^{\kappa-2} f 
+\frac{1}{2}c_4N^2(1 - \kappa)(3 - \kappa) 
  \int_0^\infty \d{r}\; r^{\kappa-4} (f - \cos f\sin f) \non
\mathop-\frac{1}{2}c_4 N^2 
  \int_0^\infty \d{r}\; r^{\kappa-2} \sin(2f) f^{'2} 
-\frac{N^2}{2}\int_0^\infty \d{r}\; r^{\kappa-2} \sin(2f)
-\int_0^\infty \d{r}\; r^\kappa \frac{\p V}{\p f} \non
= -c_4 N^2 f(0) \delta^{\kappa3},
\label{eq:babySk_nu1_mu1}
\end{align}
which is valid (converging) for $\kappa=1$, $\kappa=3$ and
$\kappa\geq 4$; the usual boundary condition for the baby Skyrmions is
$f(0)=\pi$.
This identity picks up a type-{\bf Ia} boundary charge.
A nice simple example occurs for $\kappa=1$ which is
identity \ref{id:I}+\ref{id:II}+\ref{id:VI}+\ref{id:VII}:
\begin{align}
\frac{1}{2}c_4 N^2 
  \int_0^\infty \frac{\d{r}}{r} \sin(2f) f^{'2} 
+\frac{N^2}{2}\int_0^\infty \frac{\d{r}}{r} \sin(2f)
= -\int_0^\infty \d{r}\; r \frac{\p V}{\p f},
\label{eq:babySk_nu1_mu1_kappa1}
\end{align}
while for $\kappa=3$ we have
\begin{align}
4\int_0^\infty \d{r}\; r f
-\frac{1}{2}c_4 N^2 
  \int_0^\infty \d{r}\; r \sin(2f) f^{'2} 
-\frac{N^2}{2}\int_0^\infty \d{r}\; r \sin(2f) &\non
\mathop-\int_0^\infty \d{r}\; r^3 \frac{\p V}{\p f} &= -c_4 N^2 f(0).
\label{eq:babySk_nu1_mu1_kappaeq3}
\end{align}
For the massive case there is no upper limit on $\kappa$, while for
the massless case, $\kappa$ is bounded from above as $\kappa<N+1$.

We will now consider the case of $\mu=2$, for which the
identity \ref{id:II}+\ref{id:VI}+\ref{id:VIII} reads
\begin{align}
-\left(\frac{\kappa}{2} - 1\right)
  \int_0^\infty \d{r}\; r^{\kappa-1} f^{'2} 
-\frac{\kappa c_4N^2}{2}
  \int_0^\infty \d{r}\; r^{\kappa-3} \sin^2(f) f^{'2} \non
\mathop-\frac{N^2}{2}\left(\frac{\kappa}{2} - 1\right)
  \int_0^\infty \d{r}\; r^{\kappa-3} \left(\cos(2f)-1\right) 
+\kappa\int_0^\infty \d{r}\; r^{\kappa-1} V = 0,
\end{align}
valid for $0<\kappa<2(N+1)$ for the massless case and just $\kappa>0$
for the massive case.
The baby Skyrmions, thus, have no type-{\bf II} boundary charges.
This can be understood by the fact that the baby Skyrmion is a
topological texture and not a defect. 

A particularly elegant example is given by setting $\kappa=2$,
yielding
\begin{align}
-c_4 N^2 \int_0^\infty \frac{\d{r}}{r} \sin^2(f) f^{'2} 
+2\int_0^\infty \d{r}\; r V = 0.
\label{eq:babySk_nu1_mu2_kappa2}
\end{align}
This example is particularly nice because it relates an integral of
the field to the potential energy (the last term); more precisely, it
is a virial relation relating the potential energy to the pressure
term (fourth-order derivative term).  

In some sense the above relation is a generalization of the
Pohozaev identity for baby Skyrmions.
From scaling arguments -- which Derrick's theorem is based on -- it is
intuitively clear why the kinetic term is absent; it is, as mentioned
above, classically conformal and hence does not play a role in
stabilizing the classical soliton.
This identity is a nice simple example that can be used to check the
accuracy of numerical solutions.

The last example we will consider here explicitly, is $\mu=\kappa$
which is identity \ref{id:I}+\ref{id:II}+\ref{id:VI}:
\begin{align}
c_4 N^2\left(\frac12 - \frac{1}{\kappa}\right) 
  \int_0^\infty \d{r}\; r^{\kappa-2} \sin(2f) f^{'\kappa+1}
-2c_4N^2\left(1 - \frac{1}{\kappa}\right)
  \int_0^\infty \d{r}\; r^{\kappa-3} \sin^2(f) f^{'\kappa} \non
-\frac{N^2}{2}\int_0^\infty \d{r}\; r^{\kappa-2} \sin(2f) f^{'\kappa-1}
-\int_0^\infty \d{r}\; r^\kappa \frac{\p V}{\p f} f^{'\kappa-1} = 0,
\label{eq:babySk_nu1_mueqkappa}
\end{align}
which is valid for $\kappa\geq 1$.
There is no upper limit on $\kappa$, except coming from the last term,
which in turn depends on the chosen (massless) potential.
For the massive case, there is no upper bound on $\kappa$ of course.
Let us consider the potential \eqref{eq:VbabySk_holomorphic} as an
example of a massless potential; this potential yields no upper bound
either.

\subsection{Skyrme vortices}\label{sec:skyrmevtx}

We will now consider the example of vortices in the Skyrme theory,
see \cite{Gudnason:2016yix}.
The Lagrangian is given by the Skyrme Lagrangian, which we can write
as
\begin{equation}
\mathcal{L} = -\frac{1}{2}\p_\alpha\mathbf{n}\cdot\p^\alpha\mathbf{n}
-\frac{c_4}{4}\left(\p_\alpha\mathbf{n}\cdot\p^\alpha\mathbf{n}\right)
  \big(\p_\beta\mathbf{n}\cdot\p^\beta\mathbf{n}\big)
+\frac{c_4}{4}\left(\p_\alpha\mathbf{n}\cdot\p_\beta\mathbf{n}\right)
  \big(\p^\alpha\mathbf{n}\cdot\p^\beta\mathbf{n}\big)
  -V[n^4+\i n^3],
\label{eq:LSkyrme}
\end{equation}
where $\mathbf{n}$ now is an O(4) vector field of unit length,
$\mathbf{n}\cdot\mathbf{n}=1$. 
By choosing an appropriate Ansatz for the field
\beq
\mathbf{n} = \left(
\cos f(r)\sin\chi,
-\cos f(r)\cos\chi,
\sin f(r)\sin(N\theta),
\sin f(r)\cos(N\theta)
\right),
\eeq
where $\chi$ is a constant and using the following vortex potential
\beq
V[n^4+\i n^3] = \frac{m^2}{2}\left(1 - |n^4+\i n^3|^2\right),
\eeq
the Lagrangian density reduces to
\beq
\mathcal{L} =
-\frac{1}{2}f^{'2}
-\frac{N^2}{2r^2}\sin^2 f
-\frac{c_4 N^2}{2r^2}\sin^2(f) f^{'2}
-\frac{m^2}{2}\cos^2 f,
\eeq
which is exactly the same reduced Lagrangian as for the
baby Skyrmions, see Eq.~\eqref{eq:reducedLbabySkyrmions}, except for a
different potential, i.e.,
\beq
V = \frac{m^2}{2} \cos^2 f.
\eeq

The two solitons belonging to each Lagrangian are, however,
topologically very different.
The baby Skyrmion is a texture, which means that the entire
two-dimensional space that is mapped to the target space; in
particular, the two-dimensional configuration space $\mathbb{R}^2$ is
point-compactified by identifying infinity as one point on the
manifold, such that $\mathbb{R}^2\cup\{\infty\}\simeq S^2$ is
topologically a 2-sphere. The target space is also a 2-sphere and
hence the number of textures, that is baby Skyrmions, is given by
$\pi_2(S^2)=\mathbb{Z}\ni N$.
For the Skyrme vortices, on the other hand, the solitons are defects.
This means that only the boundary of the two-dimensional space is
mapped to the target space; in particular the boundary is a circle
$S^1$ which is mapped to $S^1$ and hence the number of Skyrme vortices
is given by $\pi_1(S^1)=\mathbb{Z}\ni N$.
Note that although the full target space of the Skyrme model is $S^3$,
the construction in \cite{Gudnason:2016yix} breaks the SU(2) symmetry
explicitly down to U(1), thus providing a target space with the
topology of a circle. For more details, see \cite{Gudnason:2016yix}. 

From the point of view of the equations of motion there is no
difference of course, except for a crucial point: they obey different
boundary conditions.
In particular, the baby Skyrmion obeys $f(0)=\pi$ and $f(\infty)=0$,
whereas the Skyrme vortex obeys $f(0)=0$ and $f(\infty)=\pi/2$. 
This means that the integration by parts and the ranges of $\mu$,
$\nu$ and $\kappa$ should be reconsidered carefully.

Since the boundary conditions are different with respect to the
baby Skyrmions, we need to analyze the asymptotic behavior of $f$ for
this case of Skyrme vortices.
For small radii, we get
\beq
f = A r^N + \mathcal{O}(r^{N+2}),
\eeq
while at large radii, the linearized equation of motion for
$f=\pi/2-\delta$ reads:\footnote{We do not know the exact solution to
this differential equation, although it is linear. }
\beq
\delta''
+\frac{1}{r} \delta'
+\frac{c_4N^2}{r^2} \delta''
-\frac{c_4N^2}{r^3} \delta'
+\frac{N^2}{r^2} \delta
-m^2 \delta = 0,
\eeq
and when expanded (truncated) to leading order in $1/r$, it gives rise
to the following approximate solution 
\beq
f \simeq \frac{\pi}{2} - B K_0\left(m r\right).
\eeq
It is worthwhile to mention that this vortex is not a gauged vortex,
and thus the vortex tension diverges
logarithmically \cite{Gudnason:2016yix}.

Since the system is almost identical to that of the last subsection, 
we simply get Eq.~\eqref{eq:babySk_id} again as the main identity with
the same condition $\kappa>\mu(1-N)$.
In this case we will fix the potential to
$\frac{\p V}{\p f}=-\frac{m^2}{2}\sin(2f)$.
Setting again $\lambda=\mu$ and $\nu=1$, simplifies the identity to
identity \ref{id:II}+\ref{id:VI} as given in
Eq.~\eqref{eq:babySk_nu1}. 
The main difference between the baby Skyrmion (texture) and the
Skyrme vortex (defect) is the boundary conditions and this comes into
play when fixing the integration constants upon integrating by parts.
We see this e.g.~upon setting $\mu=1$, which is
identity \ref{id:II}+\ref{id:VI}+\ref{id:VII}: 
\begin{align}
(1 - \kappa)^2\int_0^\infty \d{r}\; r^{\kappa-2} (f-f(\infty)) &\non
\mathop+\frac{1}{2}c_4N^2(1 - \kappa)(3 - \kappa) 
  \int_0^\infty \d{r}\; r^{\kappa-4} (f - \cos f\sin f - f(\infty)) &\non
\mathop-\frac{1}{2}c_4 N^2 
  \int_0^\infty \d{r}\; r^{\kappa-2} \sin(2f) f^{'2} 
-\frac{N^2}{2}\int_0^\infty \d{r}\; r^{\kappa-2} \sin(2f) &\non
+\frac{m^2}{2}\int_0^\infty \d{r}\; r^\kappa \sin(2f)
&= c_4 N^2 f(\infty) \delta^{\kappa3},
\label{eq:Skvtx_nu1_mu1}
\end{align}
which is again valid for $\kappa=1$, $\kappa=3$ and $\kappa\geq 4$;
the boundary conditions for the vortices used in 
Ref.~\cite{Gudnason:2016yix} are $f(\infty)=\pi/2$.
This is an example of a type-{\bf Ia} boundary charge. 
Setting $\kappa=1$, yields again
Eq.~\eqref{eq:babySk_nu1_mu1_kappa1}. 
However, the case $\kappa=3$ differs from the baby Skyrmions because 
of the boundary conditions and it reads
\begin{align}
4\int_0^\infty \d{r}\; r (f-f(\infty))
-\frac{1}{2}c_4 N^2 
  \int_0^\infty \d{r}\; r \sin(2f) f^{'2} &\non
\mathop-\frac{N^2}{2}\int_0^\infty \d{r} \; r \sin(2f) 
+\frac{m^2}{2}\int_0^\infty \d{r}\; r^3 \sin(2f)
&= c_4 N^2 f(\infty).
\label{eq:skyrmevtx_nu1_mu1_kappa3}
\end{align}

We will now turn to the case of $\mu=2$, which is
identity \ref{id:II}+\ref{id:VI}+\ref{id:VIII}; again the differences
with respect to the baby Skyrmions are due to the different boundary 
conditions and the identity reads
\begin{align}
-\left(\frac{\kappa}{2} - 1\right)
  \int_0^\infty \d{r}\; r^{\kappa-1} f^{'2} 
-\frac{\kappa c_4N^2}{2}
  \int_0^\infty \d{r}\; r^{\kappa-3} \sin^2(f) f^{'2} &\non
\mathop-\frac{N^2}{2}\left(\frac{\kappa}{2} - 1\right)
  \int_0^\infty \d{r}\; r^{\kappa-3} \left(\cos(2f)+1\right) 
+\frac{\kappa m^2}{2}\int_0^\infty \d{r}\; r^{\kappa-1} \cos^2f
&= \frac{N^2}{2}\delta^{\kappa2},
\end{align}
where we have used the boundary conditions $f(0)=0$ and
$f(\infty)=\pi/2$ (only the value of $\cos(2f)$ at the boundaries
matters).
This identity picks up a type-{\bf IIa} boundary charge. 
Again, a particularly elegant example is the case of $\kappa=2$, for which
we get identity \ref{id:I}+\ref{id:II}+\ref{id:VI}+\ref{id:VIII}:
\begin{align}
-c_4N^2 \int_0^\infty \frac{\d{r}}{r} \sin^2(f) f^{'2} 
+m^2\int_0^\infty \d{r}\; r \cos^2f = \frac{N^2}{2}.
\label{eq:Skvtx_nu1_mu2_kappa2}
\end{align}
The above identity is very interesting as it is clear from the
vortices that this is indeed a generalization of the Pohozaev
identity, where the winding number is related to the potential energy.
Since the vortices at hand here enjoy a higher-derivative correction
in the form of the Skyrme term, we indeed see that the effect of the
higher derivative term is to enlarge the vortex
(``internal pressure''), which in turn requires a larger potential
energy to keep the right-hand side equal to the topological winding
number.  

Finally, the example of $\mu=\kappa$,
i.e.~identity \ref{id:I}+\ref{id:II}+\ref{id:VI} is simply given by
Eq.~\eqref{eq:babySk_nu1_mueqkappa}.

The example of the Skyrme vortex versus the baby Skyrmion is a nice
demonstration of the importance of the boundary conditions in our
class of integral identities; in particular this also demonstrates the
difference between a topological defect and a texture.
Case in point is the case of $\nu=1$, $\lambda=\kappa=\mu=2$ (identity 
\ref{id:I}+\ref{id:II}+\ref{id:VI}+\ref{id:VIII}), which is the
generalized Pohozaev identity: the sum of the two integrals
vanish in the baby Skyrmion case (see
Eq.~\eqref{eq:babySk_nu1_mu2_kappa2}) while they equal $N^2/2$ in the
Skyrme-vortex case (see Eq.~\eqref{eq:Skvtx_nu1_mu2_kappa2}).
In other words, we confirm that the boundary charges of type-{\bf IIa}
vanish at infinity for a texture due to its trivial behavior there,
while for the defect it is proportional to the winding number
squared.

\section{Examples in three dimensions}\label{sec:example3}

\subsection{Global monopoles}\label{sec:globalmonopoles}

As the first example in three dimensions, let us consider global
monopoles \cite{Vilenkin:1994} -- which are a classic example of
topological defects -- with the following Lagrangian density
\beq
\mathcal{L} =
-\frac{1}{2}\p_\alpha\boldsymbol{\phi}\cdot\p^\alpha\boldsymbol{\phi}
-\frac{\xi}{4}\left(\boldsymbol{\phi}\cdot\boldsymbol{\phi} - v^2\right)^2,
\eeq
where $\boldsymbol{\phi}$ is an O(3) vector field. 
Using the following hedgehog Ansatz
\beq
\phi^a = \frac{v f(r) x^a}{r},
\label{eq:monopole_hedgehog}
\eeq
with $a=1,2,3$, the reduced Lagrangian density reads
\beq
\mathcal{L} =
-\frac{v^2}{2}f^{'2}
-\frac{v^2}{r^2} f^2
-\frac{\xi v^4}{4}(f^2-1)^2.
\eeq
For convenience, we will now make a rescaling of the length scales,
$r\to r/v$, which will give the energy density in units of $v^4$ and
thus we arrive at 
\beq
\frac{\mathcal{L}}{v^4} =
-\frac{1}{2}f^{'2}
-\frac{1}{r^2} f^2
-\frac{\xi}{4}(f^2-1)^2.
\label{eq:Lmonopole_reduced}
\eeq
Although the reduced Lagrangian looks very similar to that giving rise
to global vortices, this defect lives in three spatial dimensions and
it is just like the global vortex only stable for a single isolated
soliton (i.e.~topological degree $N=1$).
The Ansatz \eqref{eq:monopole_hedgehog} is a hedgehog which
corresponds to a single monopole in isolation.
The generalization to higher-winding monopoles is not straightforward;
the simplest case is to assume axial symmetry and ``wind'' it $N$
times around the symmetry axis, yielding a 2-cycle of topological
degree $N$ \cite{Gudnason:2015lia}.
However, only the single monopole has spherical symmetry and thus we
can only consider that for now. 

The equation of motion is
\beq
f''
+\frac{2}{r} f'
-\frac{2}{r^2} f
-\xi(f^2-1) f = 0.
\eeq
At small $r$,
\beq
f = A r - \frac{A\xi}{10} r^3 + \mathcal{O}(r^5),
\eeq
while at large $r$, the linearized equation of motion gives the
following exact solution
\beq
f = 1 - \frac{1}{\xi r^2},
\eeq
and hence the derivative of $f$, $f'$, goes to zero asymptotically as 
\beq
f' = \frac{2}{\xi r^3}.
\eeq

Applying identity \ref{id:V} \eqref{eq:idV}, (as $G=0$), we get:
\begin{align}
2\left(1 - \frac{\kappa}{\lambda}\right) \int_0^\infty \d{r}\; r^{2\kappa-1}
  f^{'\mu}
+\left(1 - \frac{\mu}{\lambda}\right) \int_0^\infty \d{r}\;
  r^{2\kappa} f^{'\mu-1} f''
-2\int_0^\infty \d{r}\; r^{2\kappa-2} f f^{'\mu-1} \non
\mathop+\xi\int_0^\infty \d{r}\; r^{2\kappa} (1 - f^2) f^{'\mu-1} = 0,
\end{align}
valid for $\kappa>0$ and $2\kappa-3\mu<-3$.
Setting $\lambda=\mu$, we get identity \ref{id:II}+\ref{id:V}:
\begin{align}
2\left(1 - \frac{\kappa}{\mu}\right) \int_0^\infty \d{r}\; r^{2\kappa-1}
  f^{'\mu}
-2\int_0^\infty \d{r}\; r^{2\kappa-2} f f^{'\mu-1} 
+\xi\int_0^\infty \d{r}\; r^{2\kappa} (1 - f^2) f^{'\mu-1} = 0.
\end{align}
First, we can consider $\kappa=\mu$, yielding
identity \ref{id:I}+\ref{id:II}+\ref{id:V}:
\begin{align}
-2\int_0^\infty \d{r}\; r^{2\kappa-2} f f^{'\kappa-1} 
+\xi\int_0^\infty \d{r}\; r^{2\kappa} (1 - f^2) f^{'\kappa-1} = 0,
\end{align}
valid for $\kappa>3$.

Usually, the two special cases of interest are $\mu=1$ and $\mu=2$, 
due to the integration of parts yielding boundary terms.
However, the case of $\mu=1$
(i.e.~identity \ref{id:II}+\ref{id:V}+\ref{id:VI}) does not yield
convergent integrals for the single global monopole.

The case of $\mu=2$
(i.e.~identity \ref{id:II}+\ref{id:V}+\ref{id:VIII}), however, yields
interesting convergent relations:
\begin{align}
(2-\kappa) \int_0^\infty \d{r}\; r^{2\kappa-1} f^{'2}
+(2\kappa-2)\int_0^\infty \d{r}\; r^{2\kappa-3} (f^2-1) &\non
\mathop+\frac{\xi\kappa}{2}\int_0^\infty \d{r}\; r^{2\kappa-1}(f^2-1)^2 
&=\delta^{\kappa1},
\end{align}
valid for $1\leq\kappa<2$.
The right-hand side of the above identity has picked up a type-{\bf IIa}
boundary charge.
In particular, the case of $\kappa=1$ yields a very simple relation:
\beq
\int_0^\infty \d{r}\; r f^{'2}
+\frac{\xi}{2}\int_0^\infty \d{r}\; r (f^2-1)^2 
=1,
\label{eq:globalmonopole_Derrick-Pohozaev}
\eeq
where the right-hand side is the topological degree of the (single)
monopole.
This is not quite a virial relation as the integral measure is not
$r^2\d{r}$, thus the terms do not correspond to energy densities.
Nevertheless, this relation picks up a nontrivial boundary charge.

\subsection{Skyrmions}\label{sec:Skyrmions}

Now let us consider the
Skyrmion \cite{Skyrme:1961vq,Skyrme:1962vh,Manton:2004} -- being the
prime example of a texture -- which has the Lagrangian density given
in Eq.~\eqref{eq:LSkyrme}, however with $V=V[n^4]$. 
Using a hedgehog Ansatz
\beq
\mathbf{n} = \left(
\sin f(r) \sin\theta \cos\phi,
\sin f(r) \sin\theta \sin\phi,
\sin f(r) \cos\theta,
\cos f(r)
\right)
\eeq
and choosing the standard pion mass term
\beq
V = m^2(1 - n^4),
\eeq
the Skyrme Lagrangian reduces to
\beq
\mathcal{L} =
-\frac{1}{2}f^{'2}
-\frac{1}{r^2}\sin^2 f
-\frac{c_4}{r^2}\sin^2(f) f^{'2}
-\frac{c_4}{2r^4}\sin^4f
-m^2(1-\cos f).
\eeq
From the point of view of the (radially) reduced Lagrangian, the
difference between the baby Skyrmion and the Skyrmion lies in the
fourth-order derivative term; that term for the baby Skyrmion is also
the baby-Skyrmion topological number density (its integrated value is
the topological degree of the soliton), whereas the fourth-order term
in the case of the Skyrmion is geometrically similar to a curvature
term.
The boundary conditions for the Skyrmion(s) are $f(0)=\pi N$ and 
$f(\infty)=0$. 
However, only the single Skyrmion is stable in the spherically
symmetric case and thus we will fix $N=1$ in the following.

Comparing to the Lagrangian density \eqref{eq:Lcankin}, we can
identify
\beq
F = \frac{1}{r^2}\sin^2 f + \frac{c_4}{2r^4}\sin^4 f, \qquad
G = \frac{2c_4}{r^2}\sin^2(f), \qquad
V = m^2(1 - \cos f).
\eeq
The equation of motion reads
\begin{equation}
f'' + \frac{2}{r} f'
-\frac{1}{r^2}\sin(2f)
+\frac{2c_4}{r^2} \sin^2(f)f''
+\frac{c_4}{r^2}\sin(2f) f^{'2}
-\frac{c_4}{r^4}\sin(2f)\sin^2f
-m^2\sin f = 0.
\end{equation}

For small $r$, the profile function (chiral angle function) behaves
like 
\beq
f = \pi - A r + \mathcal{O}(r^3),
\eeq
while at large $r$, $f$ tends to zero and hence we can linearize the
equation of motion, getting
\beq
f'' + \frac{2}{r}f' - \frac{2}{r^2}f - m^2 f = 0,
\eeq
which has the following solutions
\beq
f=\left\{
\begin{array}{ll}
B h_{-2}^{(1)}(\i m r), \quad & m>0,\\
B/r^2, & m=0,
\end{array}\right.
\eeq
depending on whether the mass term is turned on, viz.~$m=0$ or $m>0$
and $h_n^{(1)}$ is the spherical Hankel function of the first kind.
The spherical Hankel function of the first kind with
$n=-2$, can be written as
\beq
h_{-2}^{(1)}(\i r) = \left(\frac{1}{r} + \frac{1}{r^2}\right) \e^{-r},
\eeq
and so is exponentially suppressed.

We are now ready to apply the identity \eqref{eq:id}, which yields
\begin{align}
&2\left(1 - \frac{\kappa}{\lambda}\right) \int_0^\infty \d{r}\; r^{2\kappa-1}
  (1+G)^\nu f^{'\mu}
+\left(1 - \frac{\mu}{\lambda}\right) \int_0^\infty \d{r}\; r^{2\kappa}
  (1+G)^\nu f^{'\mu-1} f'' \non
&+2c_4\left(\frac{1}{2} - \frac{\nu}{\lambda}\right) \int_0^\infty \d{r}\;
  r^{2\kappa-2} (1+G)^{\nu-1} \sin(2f) f^{'\mu+1} \non
&-4c_4\left(1 - \frac{\nu}{\lambda}\right) \int_0^\infty \d{r}\;
  r^{2\kappa-3} (1+G)^{\nu-1} \sin^2(f) f^{'\mu} 
-\int_0^\infty \d{r}\; r^{2\kappa-2} (1+G)^{\nu-1} \sin(2f) f^{'\mu-1} \non
&-c_4\int_0^\infty \d{r}\; r^{2\kappa-4} (1+G)^{\nu-1} \sin^2(f)\sin(2f)
  f^{'\mu-1} \non
&-m^2\int_0^\infty \d{r}\; r^{2\kappa} (1+G)^{\nu-1} \sin(f) f^{'\mu-1} = 0,
\end{align}
valid for $\kappa>0$ and $\mu\geq 1$ for the massive case $m>0$, while
for the massless case, $m=0$, we have the additional condition
$2\kappa<3\mu$. 

As usual, a natural choice is to set $\lambda=\mu$ and $\nu=1$, for
which we obtain identity \ref{id:II}+\ref{id:VI}:
\begin{align}
&2\left(1 - \frac{\kappa}{\mu}\right) \int_0^\infty \d{r}\; r^{2\kappa-1}
  f^{'\mu} 
+2c_4\left(\frac{1}{2} - \frac{1}{\mu}\right) \int_0^\infty \d{r}\;
  r^{2\kappa-2} \sin(2f) f^{'\mu+1} \non
&+4c_4\frac{1-\kappa}{\mu} \int_0^\infty \d{r}\;
  r^{2\kappa-3} \sin^2(f) f^{'\mu} 
-\int_0^\infty \d{r}\; r^{2\kappa-2} \sin(2f) f^{'\mu-1} \non
&-c_4\int_0^\infty \d{r}\; r^{2\kappa-4} \sin^2(f)\sin(2f)
  f^{'\mu-1} 
-m^2\int_0^\infty \d{r}\; r^{2\kappa} \sin(f) f^{'\mu-1} = 0.
\label{eq:Skyrmion_idII+VI}
\end{align}

As usual, in our construction, we can get boundary terms by choosing
$\mu=1$ or $\mu=2$. We will start with the case of $\mu=1$, getting
identity \ref{id:II}+\ref{id:VI}+\ref{id:VII}:
\begin{align}
&-2(1-\kappa)(2\kappa-1) \int_0^\infty \d{r}\; r^{2\kappa-2} f
-c_4\int_0^\infty \d{r}\; r^{2\kappa-2} \sin(2f) f^{'2} \non
&-2c_4(1-\kappa)(2\kappa-3) \int_0^\infty \d{r}\; r^{2\kappa-4}
  \left(f - \frac{1}{2}\sin(2f)\right)
-\int_0^\infty \d{r}\; r^{2\kappa-2} \sin(2f) \non
&-c_4\int_0^\infty \d{r}\; r^{2\kappa-4} \sin^2(f)\sin(2f)
-m^2\int_0^\infty \d{r}\; r^{2\kappa} \sin(f)
= -c_4 \pi \delta^{\kappa\frac{3}{2}},
\end{align}
valid for $\kappa\geq 3/2$ or $\kappa=1$.
The right-hand side of the above identity picks up a type-{\bf Ia}
boundary charge. 
In the massless case, $m=0$, however, there is the additional
condition $\kappa<3/2$, which leaves as the only possibility
$\kappa=1$: 
\begin{align}
&-c_4\int_0^\infty \d{r}\; \sin(2f) f^{'2} 
-\int_0^\infty \d{r}\; \sin(2f) 
-c_4\int_0^\infty \frac{\d{r}}{r^2}\; \sin^2(f)\sin(2f) \non
&-m^2\int_0^\infty \d{r}\; r^2 \sin(f) = 0,
\label{eq:Sk_idII+VI+VII_kappa1}
\end{align}
valid both for $m=0$ and $m>0$ (but it is the only valid identity for
$m=0$).
When $m>0$, we have another special case of $\kappa=3/2$, yielding
\begin{align}
&2\int_0^\infty \d{r}\; r f
-c_4\int_0^\infty \d{r}\; r \sin(2f) f^{'2} 
-\int_0^\infty \d{r}\; r \sin(2f) 
-c_4\int_0^\infty \frac{\d{r}}{r}\; \sin^2(f)\sin(2f) \non
&-m^2\int_0^\infty \d{r}\; r^3 \sin(f)
= -c_4 \pi.
\label{eq:Sk_idII+VI+VII_kappa3over2}
\end{align}
This illustrates nicely that a texture can also pick up boundary
charges; in this case a type-{\bf Ia} boundary charge proportional to
the Skyrme-term coefficient.

Actually, there is another type-{\bf Ia} boundary charge, that we
missed because we integrated the third term in
Eq.~\eqref{eq:Skyrmion_idII+VI} by parts, which in turn changes the
valid range of $\kappa$ from $\kappa>0$ to $\kappa\geq 3/2$.
If we only integrate the first term in \eqref{eq:Skyrmion_idII+VI} by
parts, we can write
\begin{align}
&-2(1-\kappa)(2\kappa-1) \int_0^\infty \d{r}\; r^{2\kappa-2} f
-c_4\int_0^\infty \d{r}\; r^{2\kappa-2} \sin(2f) f^{'2} \non
&+4c_4(1-\kappa) \int_0^\infty \d{r}\; r^{2\kappa-3} \sin^2(f) f' 
-\int_0^\infty \d{r}\; r^{2\kappa-2} \sin(2f) \non
&-c_4\int_0^\infty \d{r}\; r^{2\kappa-4} \sin^2(f)\sin(2f)
-m^2\int_0^\infty \d{r}\; r^{2\kappa} \sin(f)
= \pi \delta^{\kappa\frac{1}{2}},
\end{align}
which is now valid for $\kappa\geq 1/2$ for the massive case $m>0$ and in
the massless case $\kappa<3/2$ as before.
The special case is now $\kappa=1/2$, i.e.,
\begin{align}
&-c_4\int_0^\infty \frac{\d{r}}{r}\; \sin(2f) f^{'2}
+2c_4 \int_0^\infty \frac{\d{r}}{r^2}\; \sin^2(f) f' 
-\int_0^\infty \frac{\d{r}}{r}\; \sin(2f) \non
&-c_4\int_0^\infty \frac{\d{r}}{r^3}\; \sin^2(f)\sin(2f)
-m^2\int_0^\infty \d{r}\; r \sin(f)
= \pi,
\label{eq:Sk_idII+VI+VII_kappa1over2}
\end{align}
which picks up a different type-{\bf Ia} boundary charge which is
due to the kinetic term instead of the Skyrme-term coefficient.

The next interesting case, is $\mu=2$ which yields the
identity \ref{id:II}+\ref{id:VI}+\ref{id:VIII}:
\begin{align}
&2\left(1 - \frac{\kappa}{2}\right) \int_0^\infty \d{r}\; r^{2\kappa-1}
  f^{'2} 
+2c_4(1-\kappa) \int_0^\infty \d{r}\;
  r^{2\kappa-3} \sin^2(f) f^{'2} \non
&+2(\kappa-1)\int_0^\infty \d{r}\; r^{2\kappa-3} \sin^2f
+\frac{c_4}{2}(2\kappa-4)\int_0^\infty \d{r}\; r^{2\kappa-5} \sin^4 f \non
&-2m^2 \kappa \int_0^\infty \d{r}\; r^{2\kappa-1} \left(\cos f - 1\right) = 0,
\end{align}
valid for $\kappa>0$ in the massive case ($m>0$), while for the
massless case we have the additional condition $\kappa<3$.  
This case does not pick up any boundary charges.
Two particularly simple cases of the above identity are $\kappa=1$:
\begin{align}
&\int_0^\infty \d{r}\; r f^{'2} 
-c_4\int_0^\infty \frac{\d{r}}{r^3}\; \sin^4 f 
-2m^2 \int_0^\infty \d{r}\; r \left(\cos f - 1\right) = 0,
\label{eq:Sk_idII+VI+VIII_kappa1}
\end{align}
and $\kappa=2$:
\begin{align}
&-2c_4 \int_0^\infty \d{r}\; r \sin^2(f) f^{'2} 
+2\int_0^\infty \d{r}\; r \sin^2 f
-4m^2 \int_0^\infty \d{r}\; r^{3} \left(\cos f - 1\right) = 0.
\label{eq:Sk_idII+VI+VIII_kappa2}
\end{align}
None of them provide the right weight of $r$ to yield the potential
energy (the integrated value of the mass term), that is found by
setting $\kappa=3/2$:
\begin{align}
&\frac{1}{2} \int_0^\infty \d{r}\; r^2 f^{'2} 
-c_4 \int_0^\infty \d{r}\; \sin^2(f) f^{'2} 
+\int_0^\infty \d{r}\; \sin^2f
-\frac{c_4}{2}\int_0^\infty \frac{\d{r}}{r^2}\; \sin^4 f \non
&+3m^2 \int_0^\infty \d{r}\; r^2 \left(1 - \cos f\right) = 0.
\label{eq:Sk_idII+VI+VIII_kappa3over2}
\end{align}
The latter is a generalization of the Pohozaev identity
applied to Skyrmions.
Interestingly, it does provide the integrated potential energy
density, i.e.~the integral over the mass term with the standard volume
measure.
Because the Skyrmion is a topological texture as opposed to a
topological defect, the right-hand side is not the topological degree,
but just zero.

Note, that the above identity, in fact, is the volume integral over
the total energy density, but with the signs flipped on the second
and the fourth term.

It is somewhat interesting that the angular part of the kinetic term
(the third term) comes with a plus sign, whereas the radial derivative
part of the Skyrme term (second term) comes with a minus sign.
Therefore, the above relation is not just the Legendre transform in
the radial direction of the total energy density.

In fact, the Derrick scaling is evident from the above relation; the
kinetic term and the mass term both come with the same sign and are
only countered by the Skyrme term, which comes with the opposite
sign. It is thus clear that without the Skyrme term, the solution has
to vanish (the same conclusion is drawn, of course, from Derrick's
theorem). 

The last example, we will consider for the Skyrmion, is the case of
$\kappa=\mu$, i.e.~identity \ref{id:I}+\ref{id:II}+\ref{id:VI}:
\begin{align}
&2c_4\left(\frac{1}{2} - \frac{1}{\kappa}\right) \int_0^\infty \d{r}\;
  r^{2\kappa-2} \sin(2f) f^{'\kappa+1} 
-4c_4\left(1 - \frac{1}{\kappa}\right) \int_0^\infty \d{r}\;
  r^{2\kappa-3} \sin^2(f) f^{'\kappa} \non
&-\int_0^\infty \d{r}\; r^{2\kappa-2} \sin(2f) f^{'\kappa-1} 
-c_4\int_0^\infty \d{r}\; r^{2\kappa-4} \sin^2(f)\sin(2f) f^{'\kappa-1} \non
&-m^2\int_0^\infty \d{r}\; r^{2\kappa} \sin(f) f^{'\kappa-1} = 0.
\label{eq:Skyrmion_idI+II+VI}
\end{align}

\section{Discussion and conclusions}\label{sec:discussion}

In this paper, we have put forward a framework to work out families of
integral identities based on the equation of motion of a single field
profile function of a soliton solution. 
The main results are the derivation of \emph{five} boundary charges
that are constants which can depend on the Lagrangian parameters, the
boundary conditions or even in principle on the solution parameters
(characteristics of the solution); in particular the winding number.

Although we have derived 5 boundary charges, we have only been able to
realize 4 of them in convergent integral identities based on known
soliton systems. It would be interesting to either find an example of
the last boundary charge or prove that it cannot be realized (although
that seems a bit unlikely).

Another very interesting future direction was mentioned in the
introduction; namely, investigating if there is a deeper topological
origin in the boundary charges. Considering the mathematical
definition of the boundary charges as topological invariants would
definitely be interesting, but we will leave it for future work. 

For the BPS vortices, we find infinitely many integrals that are
simply related to the topological winding number (see
Eqs.~\eqref{eq:ANO_BPS_master_id} and \eqref{eq:CS_BPS_master_id}).
Although the solutions of these vortices on flat space are unknown;
in principle, we have indirectly found the solutions with these
relations; that is, we have infinitely many relations and thus enough
to determine some representation of the solutions, for any winding
number.
In practice, however, we do not -- at the present time -- know how to
neatly organize all this information in a definition of the exact
analytic profile function, that we would call the exact solution. 

The five boundary charges are not a fundamental number of boundary
charges; as should be clear from the construction, more boundary
charges can in principle be constructed by generalizing the Lagrangian
class of systems.

In all the examples we considered in this paper, the solitons have
support over the entire space, $\mathbb{R}^d$. 
In the case the soliton solutions have compact support, the identities
can still be applied by changing the integration range, but special
care may need to be taken for the boundary. 

As for generalizations, obvious ideas to generalize our framework is
to extend the single equation of motion case to systems of equations.
Such generalization should be straightforward and certainly reveal new 
interesting properties. 

Another and possibly more interesting generalization of our framework
of identities would be to consider partial differential equations
instead of ordinary differential equations. This relaxation of the
spherical (or axial) symmetry, would make the identities much more
powerful for real-world soliton calculations.
This is because the ODEs are usually easier to handle than the PDEs
and as we have mentioned at several places in the examples, many
soliton systems do not possess energetically preferred (stable)
solutions for winding numbers (topological degrees) higher than one. 
Obviously, the multi-soliton solutions are of much interest and are
generically a very complicated topic of numerical research.
It is clear that the total derivative in 2 or 3 dimensions will also,
due to Stokes' theorem, give rise to what we denoted as boundary
charges in this paper.
Therefore the generalization to higher dimensions (to PDEs) should be
possible, albeit more complicated of course.

Although we considered only flat spaces, $\mathbb{R}^d$, it should be
straightforward to generalize the identities to curved spaces. In case
the curved spaces become compact, the comment above applies; one can
still use the identities by taking the compact region of support into
account when performing the integrations.

In all the examples we considered in this paper, we only considered
cases giving convergent integrals. It may be possible to extend our
framework of identities outside the range of where the integrals
converge by introducing regularizations. Such attempt should be 
treated with care and checked carefully.

The class of theories that we considered is based on quite generic
functionals of the field profile, but with a standard kinetic term.
It is possible to generalize the standard kinetic term to e.g.~a
Dirac-Born-Infeld kinetic term. In some preliminary calculations, we
have already found that such case can easily be constructed as well. 

Finally, another possible generalization of our identities, that may
find powerful applications for studies using the AdS/CFT
correspondence, is to apply the integral identities to gravitational
systems.
This can still be restricted to cases of ODEs, but certainly needs the
extension to systems of equations, rather than single equation
identities. 
What may be discovered in such cases could be very interesting.

\subsection*{Acknowledgments}

We thank Ken Konishi for discussion.
The work of S.~B.~G.~is supported by the National Natural Science
Foundation of China (Grant No.~11675223).
The work of Z.~G.~is supported by the National Natural Science
Foundation of China (Grant No.~U1504102).
Y.~Y.~was partially supported by National Natural Science Foundation
of China (Grant No.~11471100).

\appendix

\section{Some numerical checks}\label{app:numerical_checks}

As an illustration of the identities, we perform some numerical checks
with Mathematica using its built-in {\tt NDSolve} routine.
The results are shown in the tables below.
For the calculations, we have set $m=1$, $c_4=1$, $\xi=1$.
The discrepancies between the left- and right-hand sides are displayed
with red colors for convenience. 

\begin{center}
\begin{tabular}{ll|ll}
\multicolumn{4}{c}{\emph{Abelian-Higgs BPS vortices}, Sec.~\ref{sec:AHBPSv}}\\
\hline\hline
Eq. & $N$ & LHS & RHS\\
\hline
\eqref{eq:BPSvortexN} & 1         & 1 & 1.00000000\red{452}\\
\eqref{eq:BPSvortexN} & 2         & 2 & 1.999999999\red{14}\\
\eqref{eq:BPSvortexNsq} & 1       & 1 & 1.0000000\red{4983}\\
\eqref{eq:BPSvortexNsq} & 2       & 4 & 3.999999999\red{16}\\
\eqref{eq:BPSvortexNsqplus2N} & 1 & 3 & 3.0000000\red{5818}\\
\eqref{eq:BPSvortexNsqplus2N} & 2 & 8 & 7.99999999\red{734}\\
\eqref{eq:BPSvortexresummed} ($\beta=1/2$)  & 1 & 1 & 1.0000000\red{8708}\\
\eqref{eq:BPSvortexresummed} ($\beta=1/4$)  & 2 & 1 & 0.999999999\red{72}
\end{tabular}
\end{center}

\begin{center}
\begin{tabular}{ll|ll}
\multicolumn{4}{c}{\emph{Abelian Chern-Simons-Higgs BPS vortices}, Sec.~\ref{sec:ACSHBPSv}}\\
\hline\hline
Eq. & $N$ & LHS & RHS\\
\hline
\eqref{eq:BPSAbelianCSvortexN} & 1 & 1 & 1.000000000\red{24}\\
\eqref{eq:BPSAbelianCSvortexN} & 2 & 2 & 1.9999999\red{8585}\\ 
\eqref{eq:BPSAbelianCSvortexNsq} & 1 & 1 & 1.00000000\red{296}\\
\eqref{eq:BPSAbelianCSvortexNsq} & 2 & 4 & 4.000000000\red{11}\\ 
\eqref{eq:BPSAbelianCSvortexNsqplusN} & 1 & 2 & 2.00000000\red{320}\\
\eqref{eq:BPSAbelianCSvortexNsqplusN} & 2 & 6 & 5.9999999\red{8595}\\
\eqref{eq:BPSAbelianCSvortexNsqplus2N} & 1 & 3 & 3.00000000\red{344}\\
\eqref{eq:BPSAbelianCSvortexNsqplus2N} & 2 & 8 & 7.9999999\red{7179}
\end{tabular}
\end{center}

\begin{center}
\begin{tabular}{ll|ll}
\multicolumn{4}{c}{\emph{Global Abelian vortices}, Sec.~\ref{sec:globalvtx}}\\ 
\hline\hline
Eq. & $N$ & LHS & RHS\\
\hline
\eqref{eq:globalvtx_Derrick-Pohozaev} & 1 & 1 & 0.999999\red{63}\\
\eqref{eq:globalvtx_Derrick-Pohozaev} & 2 & 4 & 3.99999\red{114}\\
\eqref{eq:global_vortex_identity1_IV+VIa} & 2 & 0.24999\red{815} & $1/4$\\
\eqref{eq:globalvtx_idII+V+VI} ($\kappa=1/2$) & 1 & 2.1\red{1781} & 2.1\red{0217}\\
\eqref{eq:globalvtx_idII+V+VI} ($\kappa=0$)   & 2 & 2.00\red{514} & 2.00\red{412}
\end{tabular}
\end{center}

\begin{center}
\begin{tabular}{ll|ll}
\multicolumn{4}{c}{\emph{Baby Skyrmions}, Sec.~\ref{sec:babyskyrmions}}\\ 
\hline\hline
Eq. & $N$ & LHS & RHS\\
\hline
\eqref{eq:babySk_nu1_mu1_kappa1} & 1   & $-2.649\red{19449617}$ & $-2.649\red{23747115}$\\
\eqref{eq:babySk_nu1_mu1_kappa1} & 2   & $-3.851600\red{40134}$ & $-3.851600\red{38448}$\\
\eqref{eq:babySk_nu1_mu1_kappaeq3} & 1 & $-3.1415926535\red{5}$ & $-3.1415926535\red{9}=-\pi$\\
\eqref{eq:babySk_nu1_mu1_kappaeq3} & 2 & $-12.56637061\red{54}$ & $-12.56637061\red{44}=-4\pi$\\
\eqref{eq:babySk_nu1_mu2_kappa2} & 1 & $\red{1.88937\times 10^{-8}}$ & 0\\
\eqref{eq:babySk_nu1_mu2_kappa2} & 2 & $\red{2.14248\times 10^{-9}}$ & 0\\
\end{tabular}
\end{center}

\begin{center}
\begin{tabular}{ll|ll}
\multicolumn{4}{c}{\emph{Skyrme vortices}, Sec.~\ref{sec:skyrmevtx}}\\ 
\hline\hline
Eq. & $N$ & LHS & RHS\\
\hline
\eqref{eq:skyrmevtx_nu1_mu1_kappa3} & 1 & 1.57079\red{984771} & $1.57079\red{632679}=\pi/2$\\
\eqref{eq:skyrmevtx_nu1_mu1_kappa3} & 2 & 6.283185\red{98699} & $6.283185\red{30718}=2\pi$\\
\eqref{eq:Skvtx_nu1_mu2_kappa2}     & 1 & 0.50000000\red{163} & $1/2$\\
\eqref{eq:Skvtx_nu1_mu2_kappa2}     & 2 & 2.00000000\red{244} & 2
\end{tabular}
\end{center}

\begin{center}
\begin{tabular}{ll|ll}
\multicolumn{4}{c}{\emph{Global monopoles}, Sec.~\ref{sec:globalmonopoles}}\\ 
\hline\hline
Eq. & $N$ & LHS & RHS\\
\hline
\eqref{eq:globalmonopole_Derrick-Pohozaev} & 1 & 0.999999\red{069} & 1
\end{tabular}
\end{center}

\begin{center}
\begin{tabular}{ll|ll}
\multicolumn{4}{c}{\emph{Skyrmions}, Sec.~\ref{sec:Skyrmions}}\\ 
\hline\hline
Eq. & $N$ & LHS & RHS\\
\hline
\eqref{eq:Sk_idII+VI+VII_kappa1} & 1 & $\red{-6.79267\times10^{-8}}$ & 0\\
\eqref{eq:Sk_idII+VI+VII_kappa3over2} & 1 & $-3.1415926\red{2093}$ & $-3.1415926\red{5359}=-\pi$\\
\eqref{eq:Sk_idII+VI+VII_kappa1over2} & 1 & 3.141\red{1627552} & $3.141\red{59265359}=\pi$\\
\eqref{eq:Sk_idII+VI+VIII_kappa1} & 1 & $\red{1.91595\times 10^{-7}}$ & 0\\
\eqref{eq:Sk_idII+VI+VIII_kappa2} & 1 & $\red{-1.24723\times 10^{-10}}$ & 0\\
\eqref{eq:Sk_idII+VI+VIII_kappa3over2} & 1 & $\red{-3.17547\times 10^{-10}}$ & 0
\end{tabular}
\end{center}

\end{document}